\documentclass[aps,twocolumn,epsfig,graphics,showpacs,floatfix,mathbbm]{revtex4}

\usepackage{amsmath,amsfonts,amssymb,graphics,graphicx,epsfig,color,times,bbm}
\usepackage[sort&compress]{natbib}
\bibpunct{[}{]}{,}{u}{}{}

\begin{document}

\bibliographystyle{apsrev}


\newcommand{\proofend}{\hfill\fbox\\\medskip }
\newtheorem{proposition}{Proposition}
\newcommand{\proof}[1]{{\bf Proof.} #1 $\proofend$}


\title{Multi-party entanglement in graph states}

\author{M. Hein$^{1,2}$, J.\ Eisert$^{3,4}$, and H.J. Briegel$^{1,2,5}$}

\address{$^1$
Theoretische Physik, Ludwig-Maximilians-Universit{\"a}t,
Theresienstra{\ss}e 37, D-80333 M{\"u}nchen, Germany}

\address{$^2$ Institut f{\"u}r Theoretische Physik, Universit{\"a}t Innsbruck,
Technikerstra{\ss}e 25, A-6020 Innsbruck, Austria}

\address{$^3$ Institut f{\"u}r Physik, Universit{\"a}t Potsdam,
Am Neuen Palais 10, D-14469 Potsdam, Germany}

\address{$^4$ Blackett Laboratory, Imperial College London,
Prince Consort Road, London SW7 2BW, UK}

\address{$^5$ Institute for Quantum Optics and Quantum Information of the Austrian Academy of Sciences,
A-6020 Innsbruck, Austria}

\date{\today}

\begin{abstract}
Graph states are multi-particle entangled states that correspond
to mathematical graphs, where the vertices of the graph take the
role of quantum spin systems and edges represent Ising interactions.
They are many-body spin states of distributed quantum systems that
play a significant role in quantum error correction, multi-party
quantum communication, and quantum computation within the framework
of the one-way quantum computer. We characterize and quantify the
genuine multi-particle entanglement of such graph states in terms
of the Schmidt measure, to which  we provide upper and lower
bounds in graph theoretical terms. Several examples and classes of
graphs will be discussed, where these bounds coincide. These
examples include trees, cluster states of different dimensions,
graphs that occur in quantum error correction, such as the
concatenated [7,1,3]-CSS code, and a graph associated with the
quantum Fourier transform in the one-way computer.
We also present  general transformation rules for
graphs when local Pauli measurements are
applied, and give criteria for the equivalence of
two graphs up to local unitary transformations, employing 
the stabilizer formalism.
For graphs of up to seven vertices we provide complete characterization
modulo local unitary transformations and graph isomorphisms.
\end{abstract}


\maketitle


\section{Introduction}

In multipartite quantum systems one can in many cases identify
constituents that directly interact with each other, whereas other
interactions play a minor role and can largely be neglected. For
example, next-neighbor interactions in coupled systems are often
by far dominant. Such quantum systems may be represented by a
graph \cite{Graph,Graph2}, where the vertices correspond to the
physical systems, and the edges represent interactions. The
concept of a graph state -- which abstracts from the actual
realization in a physical system -- is based on this intuition.

A graph state, as it is used in this paper, is a special pure multi-party
quantum state of a distributed quantum system.  It corresponds to
a graph in that each edge represents an Ising interaction between
pairs of quantum spin systems or qubits
\cite{raussen03,schlinge03,Wolfgang,Grassl}. Special instances of
graph states are codewords of various quantum error correcting codes
\cite{schlinge01}, which are of central importance when protecting
quantum states against decoherence in quantum computation
\cite{Gottesman}. Other examples are multi-party GHZ states
with applications in quantum communication, or cluster states of
arbitrary dimensions, which are known to serve as a universal
resource for quantum computation in the one-way quantum computer
\cite{Cluster,OneWay}. Yet, not only the cluster state itself is a
graph state. But also the pure state that is obtained from this
universal resource after the appropriate steps have been taken to
implement operations taken from the Clifford group. This resource
is then no longer universal, but the specific resource for a
particular quantum computation \cite{raussen03}.

In this paper we address the issue of quantifying and
characterizing the entanglement of these multi-particle entangled
states of arbitrary number of constituents. The aim is to apply
the quantitative theory of multi-particle entanglement to the
study of correlations in graph states. The underlying measure of
entanglement is taken to be the Schmidt measure \cite{Schmidt},
which is a proper multi-particle entanglement monotone that is
tailored to the characterization of such states. As holds true for
any known measure of multi-particle entanglement, its computation
is exceedingly difficult for general  states, yet for graph states
this task becomes feasible to a very high extent. 
We start by presenting general transformation rules
of graphs when local Pauli measurements are
applied locally on physical systems represented by vertices.
We present 
various upper and lower bounds for the Schmidt measure in graph
theoretical terms, which largely draw from the stabilizer theory.
These bounds allow for an evaluation of the Schmidt measure for a
large number of graphs of practical importance. We discuss these rules
 for the class of $2$-colorable graphs, which is of special practical
 importance in the context of entanglement purification \cite{Wolfgang}.
 For this class we give bounds to the Schmidt measure, that are particularly
 easy to compute.  
Moreover, we provide criteria for the equivalence of 
graph states under local unitary transformations
entirely on the level of the underlying graphs.
Finally, we present several examples, including trees, cluster states, states
that occur in the context of quantum error correction, such as the
CSS code, and the graph that is used to realize the Quantum Fourier Transformation (QFT) on three
qubits in the one-way quantum computer. The vision behind this is
to flesh out the notion of entanglement as an algorithmic
resource, as it has been put forward in Ref.~\cite{raussen03}.

The paper is structured as follows. We start by introducing
the notion of graph states of multi-qubit systems: we set the
notation concerning graph theoretical terms, and proceed by showing
how graph states are in correspondence
to graphs. Then, we recapitulate relevant
properties of the Schmidt measure as a measure of multi-particle
entanglement.
In Sec.~\ref{General} we then state general upper and
lower bounds that are formulated in the language of
graph theory.
 We also investigate the equivalence class for connected graphs up to seven
 vertices under local unitaries and graph isomorphisms.
These statements are the main results of the paper.
They are proved in  Sec.~\ref{Proofs}. We
proceed by discussing the above mentioned examples, where we use
the developed methods. Finally, we summarize what has been
achieved, and sketch further interesting steps of future research.

This paper is concerned with entanglement in multi-particle
distributed quantum systems, with some resemblance to Refs.\
\cite{Nielsen,Osterloh,Frank,Rings,Latorre,Auden,Wolf,Hideo}.
However, here we are less interested in the connection between
quantum correlations and quantum phase transition, but rather in
the entanglement of graph states that have definite applications in
quantum information theory. Entangled states associated with
graphs have also been studied in Refs.\
\cite{Rings,Buzek,Zanardi,Parker}, where bounds on shared
bipartite entanglement in multipartite quantum systems have been
studied, in order to find general rules for sharing of
entanglement in a multipartite setting.  It should, however,
 be noted that the way in which we represent entangled
states by mathematical graphs is entirely different from the way
this is done in Refs.~\cite{Rings,Buzek,Zanardi,Parker}.
Furthermore, in the present paper, we are not only concerned with
bipartite entanglement between two constituents or two groups of
constituents, but with multi-particle entanglement between many
constituents. In turn, the interaction that gives rise to the
entangled graph states is more specific, namely the one
corresponding to an Ising interaction.
Finally, as discussed above, graph states provide an
interesting class of  genuine multipartite entangled states that are
relatively easy to survey even in the regime of many parties.
Since the graph essentially encodes a preparation procedure of
the state, we will mainly examine the question of
how the entanglement in a graph state is related to the
topology of its underlying graph.

\section{Graphs, graph states, and the Schmidt measure}

\subsection{Graphs}
At the basis of our analysis lies the concept of a graph
\cite{Graph,Graph2}.
A graph is a collection of vertices
and a description of which vertices are connected by an edge.
Each graph can be represented by a diagram in a plane, where each
vertex is represented by a point and
each edge by an arc joining two not necessarily distinct vertices.
In this pictorial representation many concepts related to graphs can
be
visualized in a transparent manner. In the context of the present
paper, physical
systems will take the role of vertices,
whereas edges represent an interaction.

Formally,
an (undirected, finite) {\it graph} is a pair
\begin{equation}
    G=(V,E)
\end{equation}
of a finite
set $V\subset {\mathbbm{N}}$
and a set $E\subset [V]^2$, the elements of which are subsets of
$V$ with two elements each \cite{Graph}. The elements of $V$ are
called {\it vertices},
the elements of $E$ {\it edges}.
In the following we will only consider {\it
simple} graphs, which are graphs, that contain neither loops 
(edges connecting vertices 
with itself) nor multiple edges.

When the vertices $a,b\in V$
are the endpoints of an edge,
they are referred to as being {\it adjacent}. The adjacency relation
gives rise to an
{\it adjacency matrix} $\Gamma_G$ associated with a graph.
If  $V=\{a_1,...,a_N\}$,
then $\Gamma_G$
is a symmetric
$N\times N$-matrix, with elements
\begin{equation}
    (\Gamma_G)_{ij} =
    \left\{
    \begin{array}{ll}
    1,& \text{ if $\{a_i,a_j\}\in E$,}\\
    0 & \text{otherwise}.
    \end{array}
    \right.
\end{equation}
We will make repeated use of the neighborhood of a given vertex
$a\in V$.
This {\it neighborhood} $N_a\subset V$ is defined as the set of
vertices
$b$ for which $\{a,b\}\in E$. In other words, the neighborhood is the
set of
vertices adjacent to a given vertex. A vertex $a\in V$ with an empty
neighborhood
will be called {\it isolated vertex}.

For the purpose of
later use, we will also introduce the concept of a connected graph.
An $\{a,b\}$-path is an ordered list of vertices $a=a_1,a_2,\ldots,a_{n-1},a_n=b$, 
such that for all $i$, $a_i$ and $a_{i+1}$ are adjacent.
 A {\it connected graph}
is a graph that has an $\{a,b\}$-path for any two $a,b\in V$.
Otherwise it is referred
to as {\it disconnected}.

When a vertex $a$ is deleted in a graph, together with the edges 
incident with $a$, one obtains a new graph. For a subset of vertices
$V'\subset V$
of a graph $G=(V,E)$ let us denote
with $G-V'$ the graph that is obtained from $G$ by deleting the set
$V'$ of vertices
and all edges which are incident with an element of $V'$.
In a mild abuse of notation, we will also write $G-E'$
for the graph that results from a deletion of all edges $e\in E'$,
where $E'\subset E\subset [V]^2$
is a set of edges.
For a set of
edges $F\subset [V]^2$ we will write $G+F= (V,E \cup F)$ , and
$G \Delta F= (V, E\Delta F)$, where
\begin{equation}
    E\Delta F= (E \cup F) - (E \cap
    F)
\end{equation}
is the symmetric difference
of $E$ and $F$. Note that the symmetric difference corresponds to 
the addition modulo $2$ or the componentwise XOR if the sets are considered as
 binary vectors. Moreover, with
\begin{equation}
    E(A,B)=\{\{a,b\} \in
E : a \in A, b\in B , \, a\not=b\}
\end{equation}
we denote the set of edges between
sets $A,B\subset V$ of vertices.

\subsection{Graph states}

With each graph $G=(V,E)$ we associate a graph state. A graph
state is a certain pure quantum state on a Hilbert space ${\cal
H}_V=({\mathbbm{C}}^2)^{\otimes V}$. Hence, each vertex labels 
a two-level quantum system or qubit -- a notion that can be
extended to quantum systems of finite dimension $d$
\cite{schlinge03}. To every vertex $a\in V$ of the graph $G=(V,E)$
is attached a Hermitian operator
\begin{equation}
K_G^{(a)}
     \; = \;\sigma_x^{(a)}\prod_{b\in N_a}\sigma_z^{(b)}\,.
\end{equation}
In terms of the adjacency matrix this can be expressed as
\begin{equation}
    K_G^{(a)} =\sigma_x^{(a)}\prod_{b\in V}
(\sigma_z^{(b)})^{\Gamma_{ab}}.
\end{equation}
As usual, the matrices $\sigma_x^{(a)}, \sigma_y^{(a)},
\sigma_z^{(a)}$ are the Pauli matrices, where the upper index
specifies the Hilbert space on which the operator acts.
$K_G^{(a)}$ is an observable of the qubits associated with the
vertex $a$ and all of its neighbors $b\in N_a$. The $N=|V|$
operators $\{ K_G^{(a)}\}_{a\in V }$ are independent and they
commute.

Using standard terminology of quantum mechanics, they define a complete set of commuting
observables of the system of qubits associated with the vertex set
$V$. Thus, they have a common set of
eigenvectors, the {\it graph states}
\cite{raussen03,schlinge01,Wolfgang},
which form a basis of the Hilbert space ${\cal H}_V$. For our present
purposes, it is sufficient to choose one of these eigenvectors as a
representative of all graph states associated with $G$. We denote by
$|G\rangle$ the common eigenvector of the $K_G^{(a)}$ associated with
all eigenvalues equal to unity, i.e.,
\begin{equation}
 K_G^{(a)}|G\rangle = |G\rangle
\end{equation}
for all $a\in V$. Note that any other common eigenvector of
the set $K_G^{(a)}$ with some eigenvalues being negative are obtained
from $|G\rangle$ by simply applying appropriate $\sigma_z$
transformations at those vertices $a$, for which $K_G^{(a)}$ gives
a negative eigenvalue. In the context of quantum information
theory, the finite Abelian group
\begin{equation}
S_G=\langle \{ K_G^{(a)}\}_{a\in V }\rangle
\end{equation}
generated by the set $\{ K_G^{(a)}\}_{a\in V}$ is also called the
{\it stabilizer} \cite{Gottesman} of the graph state vector
$|G\rangle$. If the number of independent operators in $S_G$ is
less than $|V|$, then the common eigenspaces are degenerate and
can, for certain graphs $G$, be used as quantum error correcting
codes, the so-called {\it graph codes} \cite{schlinge01}. In this
case $G$ also describes a certain encoding procedure.

The graph state $|G \rangle$ can also be obtained by applying a sequence of commuting unitary two-qubit
operations $U^{(a,b)}$ to the state $| + \rangle^{\otimes V} $ corresponding
to the empty graph:
\begin{equation}\label{Ising}
|G \rangle = \prod_{(a,b) \in E} U^{\{a,b\}}\; | + \rangle^{\otimes V}\, , 
\end{equation}
where $E$ denotes the set of edges in $G$.
The unitary two-qubit operation on the vertices $a,b$, which adds
or removes the edge $\{a,b\}$, is given by
\begin{equation}\label{IsingInteraction}
U^{(a,b)}=\;P^{(a)}_{z,+}\otimes {\mathbbm{1}}^{(b)} +
P^{(a)}_{z,-}\otimes \sigma_z^{(b)}={U^{(a,b)}}^\dagger
\end{equation}
and is simply a controlled $\sigma_z$ on qubits $a$ and $b$, i.e. 
\begin{equation}\nonumber
U^{(a,b)} \; \stackrel{\cdot}{=} \; \left(
\begin{array}{cccc}
1 & 0 & 0 & 0 \\
0 & 1 & 0 & 0 \\ 
0 & 0 & 1 & 0 \\  
0 & 0 & 0 & -1 
\end{array} 
\right)\, .
\end{equation}
Here,
\begin{equation}
P^{(a)}_{z,\pm} = \frac{1\pm \sigma_z^{(a)}}{2}
\end{equation}
denotes the projector onto the eigenvector $|z,\pm \rangle$ of
$\sigma_z^{(a)}$ with eigenvalue $\pm 1$ (similarly for
$\sigma_x^{(a)}$ and $\sigma_y^{(a)}$). $U^{(a,b)}$ as in Eq.~(\ref{IsingInteraction}) is the unitary two-qubit operation which
removes or adds the edges. This is easily seen by noting that for $c
\in V-\{a,b\}$, $K_G^{(c)}$ commutes with $U^{(a,b)}$, whereas
\begin{eqnarray}\nonumber
 & & U^{(a,b)}K_G^{(a)}{U^{(a,b)}}^\dagger \cr
& = &
U^{(a,b)}\, \left( P^{(a)}_{z,-}\; +\; P^{(a)}_{z,+}\sigma_z^{(b)}
\right) \;K_G^{(a)}\nonumber \\
& = & \sigma_z^{(b)}\,K_G^{(a)} \; ,
\end{eqnarray}
because of $\sigma_x  P_{z,\pm} =   P_{z,\mp} \sigma_x $.
Since  $U^{(a,b)}= U^{(b,a)}$, similarly
\begin{equation}
U^{(a,b)}K_G^{(b)}{U^{(a,b)}}^\dagger =\sigma_z^{(a)}\,K_G^{(b)}
\end{equation}
holds, so that the transformed stabilizer corresponds to a graph
$G'$, where the edge $\{a,b\}$ is added modulo $2$.
Up to the local $\sigma_z$-rotations, this corresponds to
the Ising interaction.

An equivalence relation for graphs is inherited by the corresponding
equivalence
of state vectors. We will call two graphs $G=(V,E)$ and $G'=(V,E')$
{\it LU-equivalent}, if
there exists a local unitary $U$ such that
\begin{equation}
    |G\rangle = U|G'\rangle.
\end{equation}
Locality here refers to the systems associated with vertices
of $G=(V,E)$ and $G'=(V,E')$. Note that LU-equivalence is
different from equivalence of graphs in the graph theoretical sense,
i.e., permutations of the vertices that map neighbored vertices onto neighbored vertices.

\subsection{Schmidt measure}\label{Schmidt measure}

Graph states are entangled quantum states that exhibit complex
structures of genuine
multi-particle entanglement. It is the purpose of the present paper
to characterize and quantify the entanglement present in these
states that can be represented as graphs.
Needless to say, despite considerable research effort
there is no known computable entanglement
measure that grasps all aspects of multi-particle entanglement in an
appropriate
manner, if there is any way to fill such a phrase with meaning.
Several entanglement measures for multi-particle systems
have yet been suggested
and their properties studied \cite{Schmidt,Tangle,Plenio,Meyer,Wei,Barnum}.

We will for the purposes of the present paper use a measure of
entanglement that is tailored for characterizing the degree of
entanglement present in graph states: this is the Schmidt measure as introduced in
Ref.\ \cite{Schmidt}.

Any state vector $|\psi\rangle\in {\cal H}^{(1)} \otimes ...\otimes
{\cal H}^{(N)}$
of a composite quantum system with $N$ components
can be represented as
\begin{equation}\label{SchmidtM}
    |\psi \rangle = \sum_{i=1}^R \xi_i |\psi_i^{(1)}\rangle \otimes
...\otimes |\psi_i^{(N)}\rangle,
\end{equation}
where $\xi_i\in{\mathbbm{C}}$ for $i=1,...,R$,
and $
|\psi_i^{(n)}\rangle \in {\cal H}^{(n)}$ for $n=1,...,N$.
The {\it Schmidt measure}
associated with a state vector $|\psi\rangle$
is then defined as
\begin{equation}
E_S(|\psi\rangle ) = \log_2 (r),
\end{equation}
where $r$ is
the minimal number $R$ of terms in the sum of Eq.~(\ref{SchmidtM}) over all linear 
decompositions into product states.
It can be extended to
the entire state space (and not only the extreme points)
via a convex roof extension. This paper
will merely be concerned with pure states. More specifically,
we will evaluate the Schmidt measure for graph states
only. It should be noted, however that the Schmidt measure is a
general entanglement monotone
with respect to general local operations and classical communication
(LOCC), which  typically
leave the set of graph states.

In the multipartite case it is useful to compare the
Schmidt measure according to different partitionings, where the
components $1,...,N$
are grouped into disjoint sets. Any sequence
$(A_1,...,A_N)$ of disjoint subsets $A_i \subset V$ with
$\bigcup^N_{i=1} A_i = \{ 1, ..., N\}$
will be called a {\it partition} of $V$. We
will write
\begin{equation}
    (A_1,...A_N) \leq
    (B_1,...,B_M),
\end{equation}
if $(A_1,...A_N)$ is a {\it finer partition}
than $(B_1,...,B_M)$.
which means that every $A_i$ is contained in some $B_j$.
The latter is then a {\it coarser partition} than the former.

Among the properties that are important for the rest of the paper
are the following:
\begin{itemize}
\item [(i)]
$E_S$ vanishes on product states, i.e.,
$E_S(| \psi\rangle)= 0$ is equivalent to
\begin{equation}\label{E of product states}
|\psi \rangle =
|\psi^{(1)}\rangle \otimes ...\otimes |\psi^{(N)}\rangle.
\end{equation}
\item [(ii)]
$E_S$ is non-increasing under stochastic local operations
with classical communication (SLOCC) \cite{Schmidt,Guifre}.
Let $L^{(1)},...,L^{(N)}$ be operators acting on
the Hilbert spaces ${\cal H}^{(1)},..., {\cal H}^{(N)}$
satisfying
$ (L^{(i)})^\dagger L^{(i)} \leq {\mathbbm{1}}$,
and set $ L= L^{(1)}\otimes ...\otimes L^{(N)}$, then
    $E_S( \frac{L  |\psi\rangle}{\langle \psi| L^\dagger L |\psi\rangle 
^{1/2}}) \leq E_S(|\psi\rangle)$.
This can be abbreviated
as the statement that if
\begin{equation}\label{E under SLOCC}
    |\psi\rangle \longrightarrow_{\text {SLOCC}} |\psi'\rangle
\end{equation}
then $E_S(|\psi'\rangle ) \leq E_S(|\psi\rangle )$.
Similarly \begin{equation}\label{E under LU}
    |\psi\rangle
    \longleftrightarrow_{\text {LU}} |\psi'\rangle
    \end{equation}
    implies that
$ E_S(|\psi'\rangle ) = E_S(|\psi\rangle )$
holds, where $\longleftrightarrow_{\text {LU}}$ denotes the
interconversion via local unitaries.
Moreover, for any sequence of local projective measurements that
finally completely disentangles the state vector
$| \psi \rangle$ in each of the measurement results, we obtain the
upper bound
\begin{equation}\label{Persistency}
    E_S(|\psi \rangle) \leq \log_2 (m),
\end{equation}
where $m$ is the number of
measurement results with non-zero probability.
\item [(iii)]
$E_S$ is non-increasing under
a coarse graining
of the partitioning. If two components are merged in order to
form a new component, then the Schmidt measure can only decrease.
If the Schmidt measure of a state vector $|\psi\rangle$
is evaluated with
respect to a partitioning $(A_1,...,A_N) $, it will be
appended,
\begin{equation}
E_S^{(A_1,...,A_N)  }(|\psi\rangle),
\end{equation}
in order to avoid confusion.
The non-increasing property of the Schmidt measure then manifests
as
\begin{eqnarray} \label{E under coarse graining}
E_S^{(A_1,...,A_N)}(|\psi\rangle) \geq
E_S^{(B_1,...,B_M)}(|\psi\rangle)
\end{eqnarray}
if $(A_1,...,A_N)\leq (B_1,...,B_M)$.
For a graph $G=(V,E)$,
the partitioning where $(A_1,...,A_M)= V$ will
be referred to as {\it finest partitioning}. If no upper index is
appended to the Schmidt measure,
the finest partitioning will be implicitly assumed.

\item[(iv)]
$E_S$ is sub-additive, i.e. for the partitionings $(A_1,...,A_N)$ and $(B_1,...,B_M)$
of two different Hilbert spaces, over which $| \psi_1 \rangle$ and 
 $| \psi_2 \rangle$  are states,
\begin{eqnarray} \label{E subadditivity}
& & E_S^{(A_1,...,A_N,B_1,...,B_M)}
 \left( | \psi_1 \rangle \otimes | \psi_2 \rangle \right) \cr
& \leq & E_S^{(A_1,...,A_N)} \left( | \psi_1 \rangle \right)
+ E_S^{(B_1,...,B_M)} \left( | \psi_2 \rangle \right)\, .
\end{eqnarray}
Moreover, for any state vector $|\phi\rangle$
that is a
product state with respect to
the partitioning
$(B_1,...,B_M)$, we
have that
\begin{eqnarray} \label{E product state additivity}
& & E_S^{(A_1,...,A_N,B_1,...,B_M)}
 \left( | \psi \rangle \otimes | \phi \rangle \right) \cr
& = & E_S^{(A_1,...,A_N)} \left( | \psi \rangle \right)
\; .
\end{eqnarray}

\item[(v)] \label{E for bipartition}
For any bipartition $(A,B)$,
\begin{equation}
    E_S(|\psi\rangle)= \log_2({\text{rank}}
    (\text{tr}_A[|\psi\rangle\langle\psi|])).
\end{equation}
Moreover $E_S$ is additive within a given bipartitioning, i.e., if
$A=A_1\cup A_2$  and $B=B_1 \cup B_2$,
then
\begin{eqnarray}\label{E additivity for bipartitions}
E_S^{(A,B)}(|\psi_1\rangle \otimes |\psi_2\rangle ) & =&
E_S^{(A_1,B_1)}(|\psi_1\rangle)\nonumber\\
&+&
E_S^{(A_2,B_2)}(|\psi_2\rangle).
\end{eqnarray}
\end{itemize}

The Schmidt measure is a measure of entanglement that quantifies genuine
multi-particle entanglement. Yet, it is a coarse measure that divides pure states into classes, each of which is associated with the logarithm of a natural number or zero. 
But more detailed information can be obtained by considering more than one split
of the total quantum system.
As stated in property (ii) the Schmidt measure is a multi-particle entanglement monotone
\cite{Schmidt}.
The fact that it is a non-continuous functional on state space is a weakness when considering bipartite entanglement (where it merely reduces to the logarithm of the Schmidt rank for pure states) and in those few-partite cases where other measures are still feasible  to some extent. However, for the present purposes it turns out to be just the appropriate tool that is suitable for  characterising the multi-particle entanglement of graph states associated with potentially very many vertices.

Moreover, it should be noted that for general pure states of multipartite quantum systems the Schmidt measure is -- as any other measure of multipartite entanglement -- exceedingly difficult to compute. In order to determine the Schmidt measure $E_S$, one has to show that a given decomposition in Eq.~(\ref{SchmidtM}) with $R$ is minimal. The minimization problem involved is, as such, not even a convex optimization problem. Since $E_S$ is discrete, the minimization has to be done by ruling out that any decomposition in $R-1$ product terms exists.
According to a fixed basis $\{|0 \rangle^{(a)} ,|1 \rangle^{(a)}\}$ for each of the $N$ qubit systems, the decomposition in Eq.~(\ref{SchmidtM}) can be written as
\begin{eqnarray}
 \sum_{i=1}^R & \xi_i &  \left( \alpha_i ^{(1)} |0\rangle^{(1)} +
 \beta_i^{(1)} |1\rangle^{(1)} \right) \otimes
...\cr
& &\otimes \left( \alpha_i^{(N)} |0\rangle^{(N)}
 +  \beta_i^{(N)} |1\rangle
^{(N)}
 \right) .
\end{eqnarray}
Not taking normalization into account, which would increase the number of
 equations while decreasing the number of parameters,
Eq.~(\ref{SchmidtM}) can therefore
be rewritten as a system of nonlinear equations in the  variables
$\xi_i, \alpha_i^{(a)},\beta_i^{(a)}\in {\mathbbm{C}}$
with $i=1,...,R$ and $a=1,...,N$.
In this way one would essentially arrive at testing
 whether a system of $2^N$ polynomials in $(2N+1)\times 2^{E_S}$
complex variables has common null spaces. This illustrates that the determination
of the Schmidt measure for a general state can be a very difficult problem of
numerical analysis, which scales exponentially in the number of parties $N$
as well as in the degree of entanglement of the state itself (in terms of the
Schmidt measure $E_S$).

Remember, however, that the graph states themselves
represent already a large class of genuine multipartite entangled
states that are relatively easy to survey even in the regime of
many parties. A numerical analysis \cite{Lathauwer} seems still unrealistic in this
regime, at least until simpler procedures or generic arguments are
found. In the following, we will provide lower and upper bounds for
the Schmidt measure of graph states in graph theoretic terms,
which will coincide in many cases. Because of the complexity of
the numerical reformulation given above, we will omit the
computation of the exact value for the Schmidt measure in those
cases, where lower and upper bounds do not coincide. We will now
turn to formulating general rules that can be applied when
evaluating the Schmidt
measure on graph states for a given graph.

\section{General rules for the evaluation of the degree of
entanglement for graph states}\label{General}

In this section we will present general rules
that give rise to upper and lower bounds for the Schmidt measure,
that render the actual evaluation of the Schmidt measure feasible
in most cases. We will also present
rules that reflect local changes of the graph.
We will first merely state the bounds, the proofs can then be found
in Sec.~\ref{Proofs}.
For clarity, we will state the main
results in the form of propositions.
In Sec.~\ref{Examples} we will then apply these rules, and
calculate the Schmidt measure for a
number of graphs.

\subsection{Local Pauli measurements}\label{Pauli measurements}

It is well known that any unitary operation or projective
measurement associated with operators in the Pauli group can be
treated within the stabilizer formalism \cite{Gottesman}, and
therefore be efficiently simulated on a classical computer 
\cite{footnote-to-KGT}.
Moreover, since any stabilizer code (over a finite field) can be
written as a graphical quantum code
\cite{Grassl,schlinge02}, any measurement of operators in the
Pauli group turns a given graph state into a new one. More
precisely, consider a graph state vector $|G\rangle$  which is
stabilized by ${ S}_G = \langle \{K_G^{(a)}\}_{a \in V}\rangle $
and on which a Pauli measurement is performed. The transformed
stabilizer ${S'}$ of the new graph state vector
\begin{equation}
   U |G'\rangle = P |G\rangle
\end{equation}
after the projective
measurement associated with the projector
$P$ is up to local unitaries $U $ a stabilizer ${S}_{G'}$
according to a new graph $G'$. Here and in the following, we will
consider unit
rays corresponding to state vectors only, and for simplicity of
notation, we will write
$|\psi \rangle =   |\psi' \rangle$ for Hilbert space vectors,
if $|\psi \rangle$ and $|\psi' \rangle$ are identical
up to a scalar complex factor, disregarding normalization. We obtain
\begin{equation}
 { S'} = U  { S}_{G'} U^{\dagger}  = \langle \{U  K_G^{(a)}
U^{\dagger}\}_{a \in V}\rangle .
 \end{equation}

It will be very helpful to specify into which
graph $G$ is mapped under such a measurement, without the need of
formulating the measurement as a projection applied on Hilbert
space vectors. This is the content of the following proposition.

 Let $a \in V$ denote the vertex corresponding to the qubit
of which the observable $\sigma_z^{(a)}$, $\sigma_y^{(a)}$ or
$\sigma_x^{(a)}$ is
measured.  Corresponding to this measurement we define
unitaries $ U_{i,\pm}^{(a)}$:
\begin{eqnarray}
    U_{z,+}^{(a)}&=&{\mathbbm{1}},\,\,\,\,\,\,
     U_{z,-}^{(a)}= \prod_{b \in N_a}
    \sigma_z^{(b)},\label{uz}\\
    U_{y,+}^{(a)}  &= &\prod_{b \in N_a} \left(- i
    \sigma_z^{(b)}\right)^{1/2},\,\,\,\,\,\,
     U_{y,-}^{(a)}  = \prod_{b \in N_a} \left(i
    \sigma_z^{(b)}\right)^{1/2} \label{uy}
\end{eqnarray}
and, depending furthermore on a vertex $b_0 \in N_a$,
\begin{eqnarray} \label{uxp}
    U_{x,+}^{(a)}&=& \left(+ i \sigma_y^{(b_0)}\right)^{1/2} \prod_{b \in
N_a
-N_{b_0}-\{b_0\}}  \sigma_z^{(b)},\\
    U_{x,-}^{(a)} &=& \left(- i \sigma_y^{(b_0)}\right)^{1/2} \prod_{b \in
N_{b_0}
    -N_{a} -\{a\} }  \sigma_z^{(b)}.\label{uxm}
\end{eqnarray}

\begin{proposition}[Local Pauli measurements]\label{pauli}
    {\it Let $G=(V,E)$ be a graph, and
let $|G\rangle $ be its graph state vector. If a
measurement of $\sigma_{x}^{(a)}$, $\sigma_{y}^{(a)}$, or
$\sigma_{z}^{(a)}$ on the
qubit associated with vertex
$a\in V$ is performed, then the resulting state vector, depending on
the outcome
$\pm 1$, is given by
\begin{eqnarray}\label{ProjectiveM}
    P^{(a)}_{i,\pm} |G\rangle = |i,\pm\rangle^{(a)} \otimes U_{i,\pm}^{(a)}
|G'\rangle ,\,\,\,\,\, i=x,y,z.
\end{eqnarray}
The resulting graph is given by
\begin{eqnarray}\label{z,y measurement}
    G'=\left\{
    \begin{array}{ll}
    G-\{a\},& \text{ for $\sigma_{z}^{(a)}$,}\\
    G\Delta E(N_a,N_a) -\{a\}, & \text{ for $\sigma_{y}^{(a)}$,}\\
    \end{array}
\right.
\end{eqnarray}
and for $\sigma_{x}^{(a)}$ by 
\begin{eqnarray}
   \label{graph x measurement}
   G'&=& G \Delta \,  E(N_{b_0},N_a)
   \Delta \,  E(N_{b_0}\cap
   N_a,N_{b_0}\cap N_a) \nonumber\\
   &  &\Delta \, E(\{b_0\}, N_a - \{b_0\}) -\{a\},
\end{eqnarray}
for any $b_0 \in N_a$, if $a\in V$ is not an isolated vertex. If
$a$ is an isolated vertex, then the outcome of the
$\sigma_x^{(a)}$-measurement is $+1$, and the state is left
unchanged. }
\end{proposition}

A similar set of rules has been found independently
by Schlingemann \cite{schlinge03}.

Note that in case of a measurement of $\sigma_{y}$,
the resulting graph can 
be produced as well by simply replacing the
subgraph $G[N_a]$ by its complement $G[N_a]^c$ .
An induced {\it subgraph} $G[A]$ of a graph $G=(V,E)$
with $A\subset V$ is the graph that is obtained when
deleting all vertices but those contained in $A$, and the edges incident
to the deleted vertices.
For a measurement of $\sigma_{x}$, like the
resulting graph $G'$, the local unitary $U_{x,\pm}$
depends on the choice of $b_0$. 
But the resulting graph states arising from different choices of $b_0$ and $b'_0$ 
will be equivalent up to the local unitary $U_{b'_0}U^\dagger_{b_0}$ (see Sec.~\ref{LU classes}).
Note also that the neighborhood of $b_0$ in
$G'$ is simply that of $a$ in
$G$ (except from $b_0$).
For a sequence of local Pauli measurements,
the local unitaries have
to be taken into account, if the measured
qubit is
affected by the unitary. For the sake of completeness we
therefore summarize
the necessary commutation
relations in Table \ref{tabl},
which denote the transformation of
the measurement basis,
if a subsequent measurement
is applied to a unitarily transformed graph state.

\begin{table}
\begin{tabular}{|c|}
\hline\hline
$P_{x,\pm} \sigma_z = \sigma_z P_{x,\mp}$,\\
$P_{y,\pm} \sigma_z = \sigma_z P_{y,\mp},$\\
$P_{z,\pm} \sigma_z = \sigma_z P_{z,\pm},$ \\
\hline \hline
$P_{x,\pm} (-i \sigma_z)^{1/2} = (-i\sigma_z)^{1/2} P_{y,\mp},$ \\
$       P_{x,\pm} (i \sigma_y)^{1/2} = (i \sigma_y)^{1/2} P_{z,\pm}$,\\
$       P_{x,\pm} (- i \sigma_y)^{1/2} = (- i \sigma_y)^{1/2} P_{z,\pm}$,\\
$    P_{x,\pm} (i \sigma_z)^{1/2} = (i\sigma_z)^{1/2} P_{y,\pm},$ \\
\hline\hline
$P_{y,\pm} (-i\sigma_z)^{1/2}  = (-i\sigma_z)^{1/2} P_{x,\pm}, $ \\
$P_{y,\pm} (i\sigma_y)^{1/2}  =  (i\sigma_y)^{1/2} P_{y,\pm}, $ \\
$ P_{y,\pm} (-i\sigma_y)^{1/2}  =  (-i\sigma_y)^{1/2} P_{y,\pm},  $ \\
$ P_{y,\pm} (i \sigma_z)^{1/2}  =  (i\sigma_z)^{1/2} P_{x,\mp}, $ \\
\hline\hline
$P_{z,\pm} (-i\sigma_z)^{1/2} = (-i\sigma_z)^{1/2} P_{z,\pm},  $ \\
$P_{z,\pm} (i \sigma_y)^{1/2} = (i \sigma_y)^{1/2} P_{x,\pm},$\\
$P_{z,\pm} (- i \sigma_y)^{1/2} = (- i \sigma_y)^{1/2} P_{x,\pm},$\\
$P_{z,\pm} (i \sigma_z)^{1/2} = (i\sigma_z)^{1/2} P_{z,\pm},  $ \\
\hline\hline
\end{tabular}
\caption{ The relevant commutation relations for Pauli projections and Clifford
operators.}\label{tabl}
\end{table}

\begin{figure}[th]
\includegraphics[width=9.2cm]{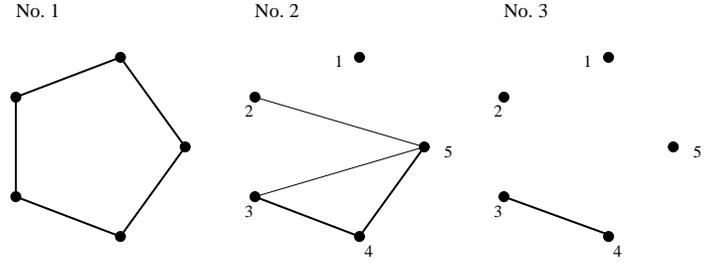}  
\caption{\label{fig:xMeasurementExample1} Example for a $\sigma_x$-measurement at vertex
$1$ in graph No.~1, which is followed by a $\sigma_z$-measurement at vertex $2$:
In graph
No.~1 a $\sigma_x$-measurement is performed at the vertex $1$.
For the application of the rule in Eq.~(\ref{graph x measurement}), vertex $2$ was chosen as
the special neighbor $b_0$, yielding graph
No.~2 up to a local unitary $U_{x,\pm}^{(1)}=  (\pm i \sigma_y^{(2)} )^{1/2}  $.
As stated in Table \ref{tabl}, the subsequent $\sigma_z$-measurement on the new graph state is therefore
essentially another $\sigma_x$-measurement, now at vertex $2$ with a single neighbor $b_0=5$.
The final graph is then graph No.~3.
}
\end{figure}

Figure~\ref{fig:xMeasurementExample1} shows two subsequent applications of
the rather complicated $\sigma_x$-measurement.
We will give a simplified version of this rule in Sec.~\ref{LU classes}.
Apart from the trivial case of a $\sigma_x$-measurement at
an isolated vertex, both measurement results $\pm1$ of a local Pauli
measurement are attained
with probability $1/2$ and yield locally equivalent graph state
vectors  $|G'\rangle$
and $|G''\rangle$. Therefore, we have
\begin{equation}\label{E under projective measurement}
    E_S(|G'\rangle ) \leq E_S(|G\rangle ) \leq E_S(|G'\rangle ) \,+\,
1\; .
\end{equation}
According to Eq.~(\ref{Persistency}), for any measurement
sequence of $\sigma_x$, $\sigma_y$ or $\sigma_z$ that yields an
empty graph, the number of local measurements in this sequence
gives an upper bound on the Schmidt measure of the corresponding
graph state. In the following we will call the minimal number of
local Pauli measurements to disentangle a graph state its {\it
Pauli persistency} (see Ref.\ \cite{Cluster}). Since each $\sigma_z$
measurement deletes all edges incident to a vertex, any subset
$V'\subseteq V$ of vertices in a graph $G$, to which any edge of
$G$ is incident, allows for a disentangling sequence of local
measurements. In graph theory those vertex subsets are called {\it
vertex covers}.

\noindent
\begin{proposition}[Upper bound via persistency]
The Schmidt measure of any graph state vector
$|G\rangle$ is bounded from above by the Pauli
persistency.
In particular, the Schmidt measure
is less than or equal to the size of the minimal
vertex cover of the corresponding graph $G$.
\label{perst}
\end{proposition}

\begin{widetext}

\begin{figure}[th]
\includegraphics[width=13cm]{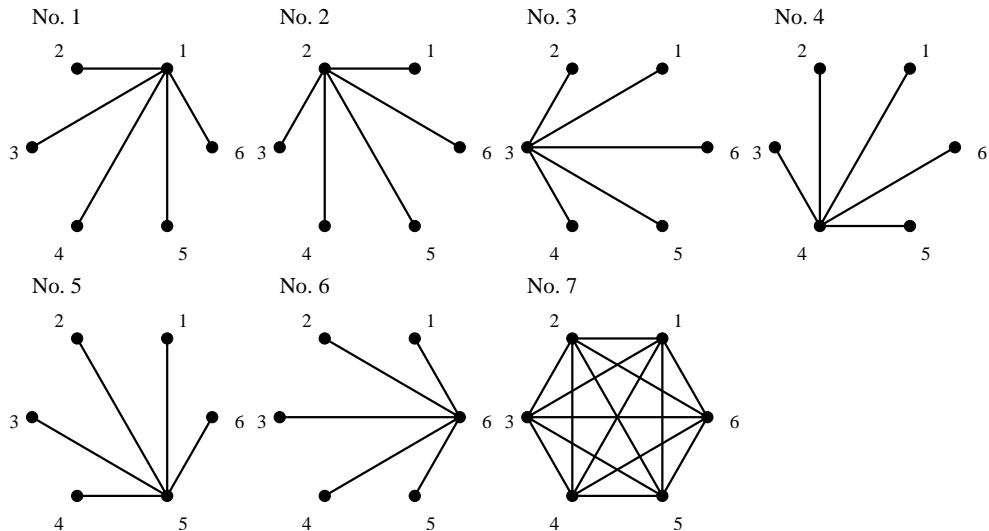}  
\caption{\label{fig:LUclassExample3} A single $\sigma_y$-measurement at an arbitrary vertex
in the complete graph No.~7 suffices to disentangle the corresponding state. Similarly, a single $\sigma_z$-measurement
at the central vertex in graphs No.~1--6 or a single $\sigma_x$-measurement at
the non-central vertices is a disentangling measurement. This is due to the fact
 that all graphs (No.~1--7) are locally equivalent by
 local unitaries, which transform
 the measurement basis correspondingly.}
\end{figure}

\end{widetext}

For graphs with many edges a combination of $\sigma_z$
and $\sigma_y$ will give better bounds than restricting to
$\sigma_{z}$ measurements only.
For example, due to Eq.~(\ref{z,y measurement}),
any complete graph (in
which all vertices are adjacent) can be disentangled by just one
$\sigma_y$-measurement
at any vertex. As we will show, this corresponds to the fact
that
these graph states are LU-equivalent to the GHZ-type graph states, in
which
every vertex is adjacent to the same central vertex (see Fig. \ref{fig:LUclassExample3}).

\subsection{Schmidt measure for bipartite splits}\label{E_S for bipartite splits}

For a bipartition $(A,B)$ of the graph $G=(V,E)$ let
$G_{AB}=(V,E_{AB})$ denote the subgraph of $G$ which is induced by
the edges  $E_{AB}\equiv E(A,B)$ between $A$ and
$B$. Moreover, $\Gamma_{AB}$ will denote the
$|A|\times|B|$-off-diagonal submatrix of the adjacency matrix
$\Gamma_G$ according to $G$, which represents the edges between
$A$ and $B$:
\begin{equation}\label{Gamma for bipartition}
\left(\begin{array}{cc}
  \Gamma_{A} & \Gamma_{AB}^T \\
   \Gamma_{AB} & \Gamma_{B} \\
\end{array}\right)
= \Gamma_G,
\end{equation}
and similarly
\begin{equation}
\left(\begin{array}{cc}
  0 & \Gamma_{AB}^T \\
 \Gamma_{AB} & 0\\
\end{array}\right)
= \Gamma_{G_{AB}}.
\end{equation}

\begin{proposition}[Bi-partitioning]\label{bip}
The partial trace with respect to any partition $A$ is
\begin{equation}\label{partial trace}
{\rm tr}_A [| G \rangle \langle G | ]= \frac{1}{2^{|A|}}
\sum\limits_{{\mathbf z} \in {\mathbbm F}_2^{A} }
 U({\mathbf z}) | G-A \rangle\langle G-A| U({\mathbf z})^\dagger
\end{equation}
where ${\mathbbm F}_2$ denotes the integer field $\{0,1\}$ with
addition and multiplication modulo $2$.
The local unitaries are defined as
\begin{eqnarray}\label{U in partial trace}
    U({\mathbf z})=
    \prod\limits_{a \in A} \left( \prod
    \limits_{b\in N_a} \sigma_z^{(b)} \right)^{z_a}.
\end{eqnarray}
Therefore, the Schmidt measure of a graph state
vector  $| G \rangle$ with respect
to an arbitrary bipartition
$(A,B)$ is given by the rank of the submatrix
$\Gamma_{AB}$ of the adjacency matrix $\Gamma_{G}$,
\begin{eqnarray}\label{Schmidt rank}
    E_S (| G \rangle ) &\geq&
E_S^{(A,B)} (| G \rangle )\nonumber\\
&= &{\rm log}_2 \left(\text{\rm
rank}
\left( \text{tr}_A [| G \rangle \langle G | ]\right) \right)
\nonumber\\
&=&  \text{\rm rank}_{{\mathbbm F}_2} (\Gamma_{AB} )= \frac{1}{2}
\text{\rm rank}_{{\mathbbm F}_2}(\Gamma_{G_{AB}}).
\end{eqnarray}
\end{proposition}

From Eq.~(\ref{partial trace}) one may as well compute that the
reduced entropy of $|G\rangle$ according to the bipartition
$(A,B)$ and the Schmidt rank coincide, if the base-2- logarithm is taken.
with respect to $2$. This simply expresses the fact that, for a non
empty graph, $|G\rangle$ is the ``maximally'' $(A,B)$-entangled
state vector with $2^{E_S^{(A,B)}}$ Schmidt coefficients.
If one maximizes over all bipartitionings
$(A,B)$ of a graph $G=(V,E)$, then according to
Eq.~(\ref{E under coarse graining})
one obtains a lower bound for the
Schmidt measure
with respect to the finest partitioning.

Note that the Schmidt rank of a graph state is closely related to
error correcting properties of a corresponding graph code. Let $A$ be
a partition, according to which $|G \rangle $ has maximal
Schmidt rank. Then, according to Ref.\ \cite{schlinge01},
choosing a subset
$X\subseteq A$, the graph code, which encodes an input on vertices
$X$ in output on vertices $Y=V-X$ according to $G$,
detects the error configuration $E=A-X$, i.e., any errors occurring on
only one half of the vertex set $E$ can be corrected.
In particular, all {\it strongly error
correcting graph codes} in Ref.~\cite{schlinge01} must have Schmidt
measure
$|V|/2$.

\begin{proposition}[Maximal Schmidt rank]\label{sufficient crit for
max rank}
A {\it sufficient criterion} for a bipartite
split $(A,B)$ to have  maximal Schmidt rank is that the graph
$G_{AB}$ contains no cycles, and that the smaller partition
 contains at most one leaf with respect to the subgraph $G_{AB}$.
If $G_{AB}$ is not connected, then it is sufficient that
the above criterion holds for every connected
component of $G_{AB}$.
 \end{proposition}

A {\it leaf} is a vertex of degree 1, i.e., a vertex to
which
exactly one edge is incident \cite{Graph}.
It is finally important to note that
the maximum Schmidt measure with respect to all bipartite partitions
is essentially  the quantity considered in Ref.~\cite{SimulClassical} in the
context of an efficient simulation of a quantum algorithm on a classical computer.
If this quantity has the appropriate asymptotic behaviour in the number $n$
of spin systems used in the computation,
then an efficient classical algorithm simulating 
the quantum dynamics can be constructed.

Note finally that, as an immediate corollary of the 
above considerations, the degree of entanglement 
depends only on the area of the boundary between distinguished
regions of regular cluster states, i.e., graph states where
in a regular cubic lattice nearest neighbors are connected by
an edge. If one considers periodic
boundary conditions, one may distinguish a
cuboid forming part $A$ 
from the rest of the graph $B$, and ask for the
bipartite entanglement. It follows immediately
that since the interior regions may be completely
disentangled, the degree of entanglement is linear
in the number of vertices forming the boundary of
the two regions. The 
corners are then counted just as one
maximally entangled pair of two-spin systems. 

\subsection{Deleting edges and vertices}

For graphs with a large number of vertices or edges, it
is useful to identify bounds for the Schmidt measure when local
changes to the graph are applied. As an example we give two rules, 
that bound the changes to the Schmidt measure
if an edge or a vertex is deleted or added.

\begin{proposition}[Edge rule]\label{edger}
By deleting or adding edges $e=\{a,b\}$
 between two vertices
$a,b \in V$ of a graph $G$ the Schmidt measure of the resulting
graph $G'=G\pm \{e\}$ can at most decrease or increase by one,
i.e.,
\begin{equation}\label{plusminus edge}
|E_S(|G'\rangle) - E_S(|G\rangle)| \leq 1.
\end{equation}
\end{proposition}

\begin{proposition}[Vertex rule]\label{vertexr}
If a vertex $a$ (including all its incident edges) is deleted, the
Schmidt measure of the resulting graph $G'=G- \{a\}$ cannot
increase and will at most decrease by one, i.e.,
\begin{equation}
    E_S(|G'\rangle)
    \leq E_S(|G\rangle)
    \leq E_S(|G'\rangle) + 1.
\end{equation}
\end{proposition}

\subsection{Bounds for $2$-colorable graphs}

Graphs may be colorable. A
proper {\it $2$-coloring of a graph} is a
labeling $V\longrightarrow  \{1,2\}$, such that all adjacent vertices
are associated with a different element from $\{1,2\}$, which can be
identified with two colors.
In graph theory these graphs are also called 'bipartite graphs',
since the set of vertices can be partitioned into two disjoint sets,
such that no two vertices within
 the same set are adjacent. It is a well known fact in graph theory that
a graph is $2$-colorable iff it does not contain any cycles of odd length.

As has been shown in Ref.~\cite{Wolfgang}, for every graph state
corresponding to a $2$-colorable graph, a multi-party entanglement
purification procedures exists: Given any $2$-colorable graph
state vector $|G\rangle$ on $|V|$ qubits, by means of LOCC
operations a general mixed state $\rho$ on $|V|$ particles can be
transformed into a mixed state, which is diagonal in a basis of
orthogonal states that are LU-equivalent to $|G\rangle$. Given
that the initial fidelity is sufficient, an ensemble of those
states then can be purified to $|G\rangle$. Thus $2$-colorable
graph states provide a reservoir of entangled states between a
large number of particles, which can be created and maintained
even in the presence of  decoherence/noise. For the class of these
graph states the lower and upper bounds to the Schmidt measure can
be applied.

\begin{proposition}[2-colorable graphs]\label{twoc}
For 2-colorable graphs $G=(V,E)$
the Schmidt measure is bounded from below
by half the rank of the adjacency matrix of the graph, i.e.,
\begin{equation}
    E_{S}(|G\rangle)\geq \frac{1}{2} {\text{rank}}_{{\mathbbm F}_2}
(\Gamma_G)
\end{equation}
and from above by the size of
the smaller partition of
the corresponding bipartition. In particular,
for a 2-colorable graph,
\begin{equation}\label{E for 2-colorable graphs}
    E_{S}(|G\rangle) \leq \lfloor\frac{|V|}{2} \rfloor.
\end{equation}
If $\Gamma_G$ is invertible, then equality holds in Eq.~(\ref{E for
2-colorable graphs}).
\end{proposition}
Note that any graph $G$,
which is not 2-colorable, can be turned into 2-colorable one $G'$
simply by deleting the appropriate vertices on cycles with odd length.
Since this corresponds to $\sigma_z$ measurements, by Eq.~(\ref{E
under projective measurement}),
\begin{eqnarray} E_S(|G\rangle) & \leq &
 E_S(|G'\rangle) \,+\, M \; \leq \; \lfloor \frac{|V - M|}{2}
\rfloor \,+\, M \cr
& \leq & \lfloor \frac{|V|+ M}{2}  \rfloor \;
,
\end{eqnarray}
 where $M$ denotes the number of removed vertices. Moreover
note that the number of induced cycles with odd length certainly
bounds $M$ from above.

We also note that whereas local $\sigma_x$- or $\sigma_z$-measurements in $2$-colorable
graphs will yield
graph states according to $2$-colorable graphs, $\sigma_y$-measurements of $2$-colorable graphs can lead to
graph states which are not even locally equivalent to $2$-colorable graphs.
It is certainly true
that a $2$-colorable graph remains $2$-colorable after application of the $\sigma_z$-measurement rule
Eq.~(\ref{z,y measurement}), since after deletion of a vertex in a
$2$-colorable graph the graph still does not contain any cycles of odd length.

Now let $G$ be a $2$-colorable graph with the bipartition $A$ of sinks and $B$ of sources,
in which the observable $\sigma_x$ is measured at vertex $a\in A$.
Then, the set $E(N_{b_0}\cap N_a,N_{b_0}\cap N_a)$ in Eq.~(\ref{graph x measurement})
is empty and $E(N_{b_0},N_a)$ only consists of edges between $A$ and  $B$.
Moreover, after adding all
edges of the last set (modulo $2$) to the edge set of the graph $G$, the measured
vertex $a\in A$, as well as its special neighbor $b_0\in B$, are isolated, so that
in the last step of adding $E(\{b_0\}, N_a - \{b_0\})$ the vertex  $b_0$ simply gets all
neighbors $N_a-\{b_0\}\subset B$ in $G$. So after application of this rule the new
graph $G'$ has the $2$-coloring with partitions  $A'=A-\{a\}\cup \{ b_0\}$ and $B'=B-\{b_0\}$.
\begin{figure}[th]
\includegraphics[width=6.5cm]{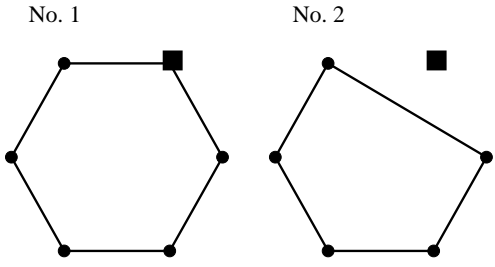}  
\caption{\label{fig:yMeasurementbipartite} Whereas graph No.~1 is $2$-colorable, the resulting
graph No.~2 after a  $\sigma_y$-measurement at the vertex
$\vrule height7pt width7pt depth0pt$ is not $2$-colorable. Also, none of the
$132$ (or $3$) representatives in the corresponding equivalence class
(if graph isomorphisms are included) is $2$-colorable.}
\end{figure}
A counterexample to a corresponding assertion for $\sigma_y$-measurements is provided in Fig.
\ref{fig:yMeasurementbipartite}. The resulting graph even has no locally
equivalent representation as a $2$-colorable graph. This is
because the corresponding equivalence class No.~8 in Table \ref{tab1} has
no $2$-colorable representative.



\begin{figure}[th]
\includegraphics[width=9cm]{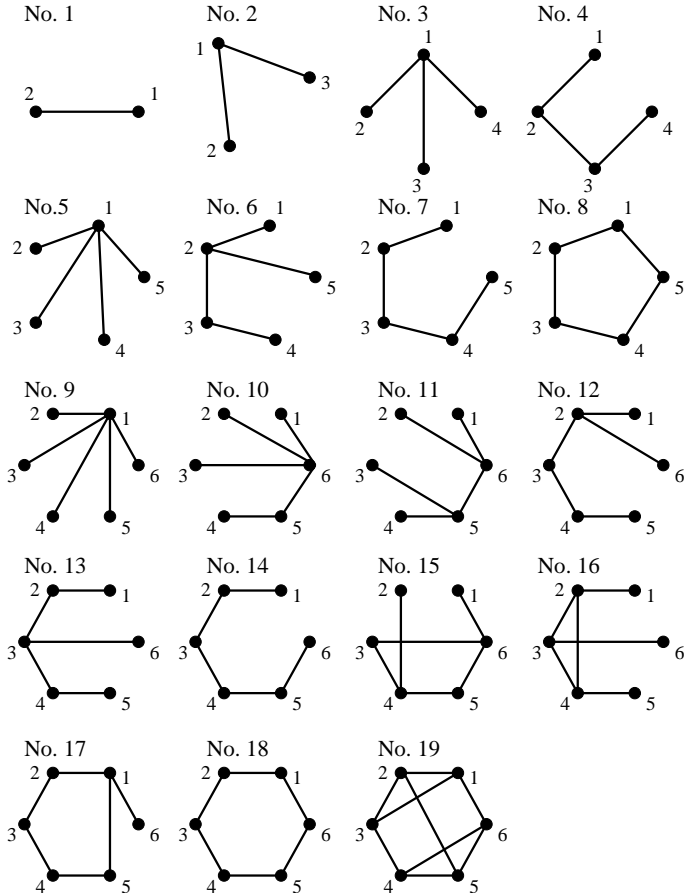}  
\caption{\label{fig:List1} List of connected graphs with up to six vertices that are not equivalent under LU transformations and graph isomorphisms.}
\end{figure}

\begin{figure}[th]
\includegraphics[width=9cm]{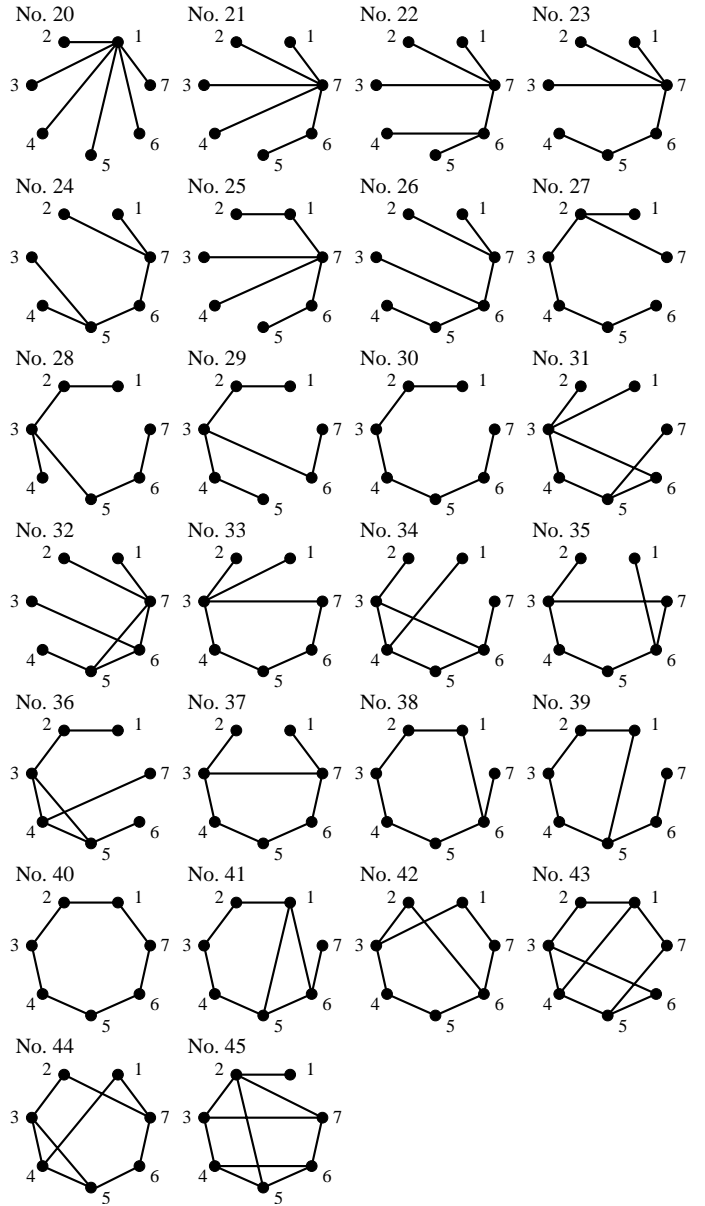}
\caption{\label{fig:List2} List of connected graphs with seven vertices that are not equivalent under LU transformations and graph isomorphisms.}
\end{figure}

\subsection{Equivalence classes of graph states under local
unitaries}\label{LU classes}

Each graph state vector $|G\rangle$
corresponds uniquely to a graph $G$. However, two graph states
can be LU-equivalent, leading to two different graphs. Needless to
say,
this equivalence relation is different from the graph isomorphisms
in graph theory. We have
examined the graph states of all non-isomorphic (connected) graphs
with up to
seven vertices \cite{Classification}. More precisely, from the set of all possible graphs
with seven vertices ($2^{\genfrac{(}{)}{0pt}{}{7}{2}} \approx 2\times 10^{6}$
possibilities), we have considered the subset of
all connected graphs on up to seven vertices which are
non-isomorphic with respect to graph isomorphisms, i.e., permutations
of the
vertices that map neighbored vertices onto neighbored vertices.
Of the $995$ isomorphism classes of corresponding graph states,
$45$ classes have turned out to be not invariant
under local unitary operations (with respect to the finest partitioning).
Moreover, within each of these classes all graph states are equivalent modulo local unitaries {\em and}
additional graph isomorphisms, which corresponds to the exchange of
particles. If we exclude the graph isomorphisms as, e.g., in quantum communication
scenarios, the number of inequivalent classes of graph states would even be
larger. In Fig. \ref{fig:List1} and \ref{fig:List2} we give a list of simple
representatives of each equivalence class. 

To test for local equivalence we have only considered local unitaries
within the corresponding local Clifford group. But by considering the
Schmidt rank with respect to all possible
bipartitions, the corresponding lists of Schmidt ranks for each
representative turned out to be different even if we allow arbitrary
permutations of the vertices. This shows that the
found sets of locally invariant graph states are maximal.

\begin{figure}[th]
\includegraphics[width=9cm]{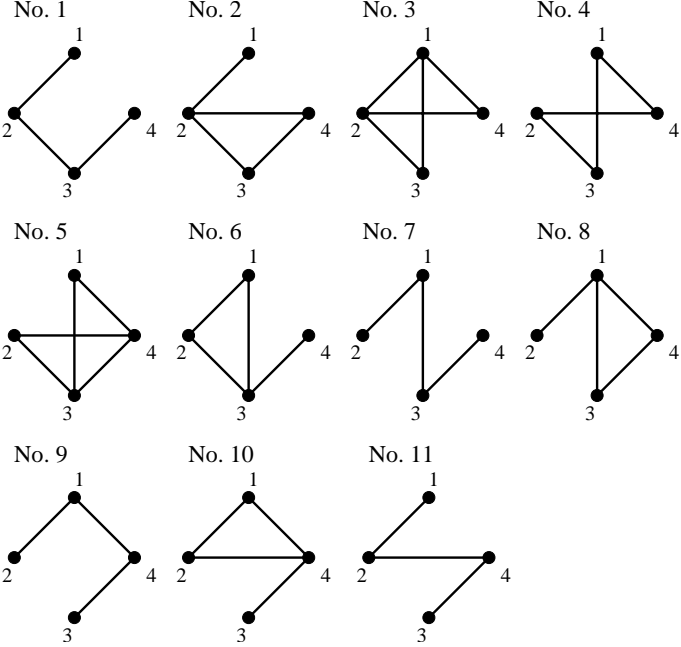}
\caption{\label{fig:LUruleExample1} An example for an successive application of the LU-rule,
which exhibits the whole equivalence class associated with graph No.~1. The rule is successively
applied to that vertex of the predecessor, which is written above the arrows of the
following diagram:}
\begin{eqnarray}  \rm{No.\ }1 \, {{3 \atop \longrightarrow} \atop} \, \rm{No.\ }2 \, {{2 \atop \longrightarrow} \atop} \, \rm{No.\ }3 \, {{3 \atop \longrightarrow}\atop} \,\rm{No.\ }4 \, {{1 \atop \longrightarrow} \atop } \,\rm{No.\ }5 \, {{3 \atop \longrightarrow} \atop } \,\rm{No.\ }6 \cr {{1 \atop \longrightarrow}\atop } \,\rm{No.\ }7 \, {{3 \atop \longrightarrow } \atop } \,\rm{No.\ }8 \, {{4 \atop \longrightarrow} \atop } \,\rm{No.\ }9 \, {{1 \atop \longrightarrow} \atop } \,\rm{No.\ }10 \, {{2 \atop \longrightarrow} \atop } \,\rm{No.\ }11 \nonumber
\end{eqnarray}
\end{figure}

Having this enormous reduction in mind, it is desirable to find
simple rules in purely graph theoretic terms, giving at least
sufficient conditions for two graph states to be equivalent by means
of local unitaries. The subsequent rule implies such a simplification:
The inversion of the subgraph $G[N_a]\mapsto G[N_a]^c$, induced by
the neighborhood $N_a$ of any vertex $a \in V$, within a given graph,
gives a LU-equivalent graph state. In graph theory this transformation 
$\tau_a:\, G\mapsto \tau_a(G)$, where the edges set $E'$ of $\tau_a(G)$ is obtained from
the edge set $E$ of $G$ by $E' = E
\Delta E(N_a,N_a)$, is known as {\it local complementation} \cite{Bouchet}.
With this notation the corresponding rule for graph states can be stated
as follows:

\begin{proposition}[LU-equivalence]\label{loc}
Let $a\in V$ be an arbitrary vertex of two graphs $G=(V,E)$, 
then
$|\tau(G)\rangle = U_a(G)\,|G\rangle$
with local unitaries of the form
\begin{eqnarray}
    U_a(G)= \left(- i \sigma_x^{(a)}\right)^{1/2}\;\prod_{b \in N_a} 
\left( i
\sigma_z^{(b)}\right)^{1/2} \propto \sqrt{K_G^{(a)}}.
\end{eqnarray}
\end{proposition}

\begin{figure}[th]

\includegraphics[width=9cm]{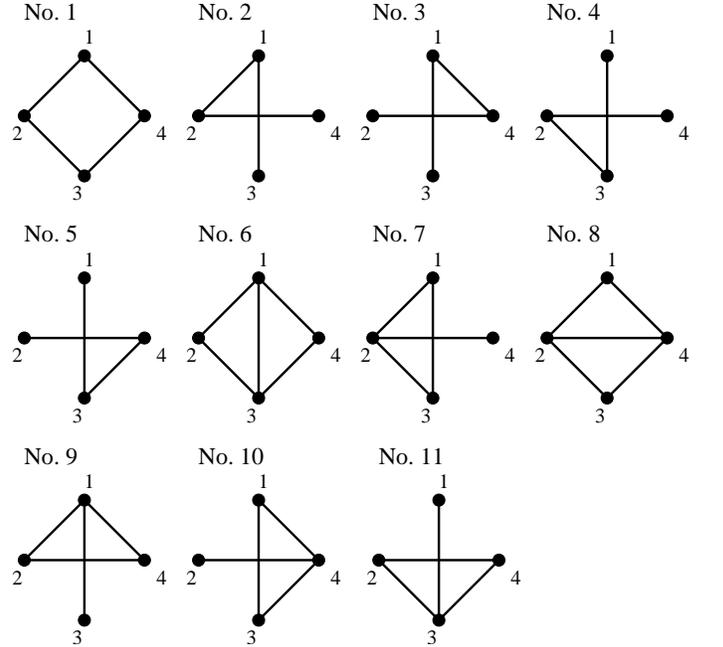}
\caption{\label{fig:LUclassExample2} An example of an equivalence class
which is a proper subset of class No.~4 in Fig.~\ref{fig:List1}}
\end{figure}

\begin{figure}[th]

\includegraphics[width=5cm]{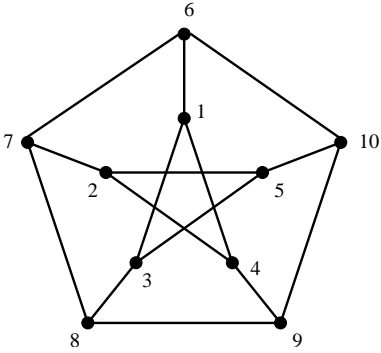} 
\caption{\label{fig:PetersenGraph} The Petersen graph. The depicted labeled graph is not LU-equivalent 
to the graph which
is obtained from it by exchanging the labels at each end of the five "spokes", i.e., the graph isomorphism which 
permutes the vertices $1,2,3,4,$ and $5$ with $6,7,8,9,$ and $10$, respectively.}
\end{figure}

This rule was independently found by Van den Nest \cite{Maarten} and Glynn \cite{Glynn02}, who showed also that a successive application of this rule suffices to generate the complete orbit of any graph state under local unitary operations within the Clifford group \cite{BouchetAlternative}. 
Figure~\ref{fig:LUruleExample1} shows an example of how to repeatedly apply this rule
in order to obtain the whole equivalence class of a graph state. Note that the set of graphs
in Fig.~\ref{fig:LUruleExample1} do not exhaust the entire class associated with graph No.~4 in Fig.~\ref{fig:List1}~.
In Fig.~\ref{fig:LUclassExample2} we show another set of graphs that is a proper subset of the class No.~4 in Fig.~\ref{fig:List1}. No graph in Fig.~\ref{fig:LUruleExample1} is locally equivalent to any graph in the equivalence class represented in Fig.~\ref{fig:LUclassExample2}, though both belong to the same equivalence class when considering both, local unitary transformations {\it and} graph isomorphisms, as depicted in Fig.~\ref{fig:List1}.
For any partition $A$ the Schmidt rank $E_S^{(A,A^c)}$ is an invariant under arbitrary local unitaries,
which is formulated in purely graph theoretic terms. 
Considering the list of Schmidt ranks with respect to all partitions, one therefore obtains a set of invariants for graphs under local complementations $\tau$, which was already considered in graph theory,  known as the {\it connectivity function } \cite{Bouchet}. For the equivalence classes in Fig.~\ref{fig:LUruleExample1} and Fig.~\ref{fig:LUclassExample2}, for example, the corresponding lists of Schmidt ranks or connectivity functions do not coincide, implying that the corresponding set of graph states are not equivalent neither under local Clifford group operations nor under general local unitaries.  We note that the Schmidt rank list does not provide a complete set of invariants that would characterize all equivalence classes under local Clifford group operations. For the Petersen graph \cite{Flaas} shown in Fig.~\ref{fig:PetersenGraph} and the isomorphic graph, which is obtained by exchanging the labels at each end of the five "spokes", no local Clifford operation exists (i.e., sequence of local complementations) that transforms one graph into the other, although the Schmidt rank lists for both graphs coincide. For a complete set of invariants the polynomial invariants in Ref.~\cite{Hans} can be considered.
For example the number of elements $I_A(G)$ in the stabilizer of the graph state vector 
$|G\rangle$, that act non-trivially exactly on 
the vertices in $A$, corresponds to 
homogeneous polynomial invariants of degree $2$. Moreover one can show that
\begin{equation}
\sum_{B\subseteq A} I_B(G) = 2^{\left(|A|-E_S^{(A,A^c)}\right)}\, .
\end{equation} 
Therefore, the list of invariants $I_A(G)$ with respect to all partitions $A$ essentially contains the same information for graph states as the Schmidt rank list.
But, as discussed in Refs.~\cite{Hans,MaartenInv}, by considering more invariants corresponding to homogeneous polynomials of different degrees one can (in principle) obtain finite and complete sets of invariants for local Clifford operations, as well as for arbitrary local unitaries. 

Finally, the LU-rule can be used to derive the $x$- and $y$-measurement rule from the simple $z$-measurement rule:
With commutation relations similar to those in Table~\ref{tabl} it is easy to see, that
$ P^{(a)}_{x,\pm} =  U_{b_0}(G) P^{(a)}_{y,\pm} U^\dagger_{b_0}(G)$ and 
$ P^{(a)}_{y,\pm} =  U_{a}(G) P^{(a)}_{z,\mp} U^\dagger_{a}(G)$ holds, where $b_0$ is a neighbor of $a$.
With the notion of local complementation at hand, we can then rewrite the resulting states in Proposition~\ref{pauli}
after the Pauli measurement in the simplified form:
\begin{eqnarray}
 P^{(a)}_{z,\pm} |G\rangle  & = &  |z,\pm\rangle^{(a)} \otimes U_{z,\pm}^{(a)} |G-a\rangle ,\\
 P^{(a)}_{y,\pm} |G\rangle  & = &  |y,\pm\rangle^{(a)} \otimes U_{y,\pm}^{(a)} |\tau_a(G)-a\rangle, \\
 P^{(a)}_{x,\pm} |G\rangle  & = &  |x,\pm\rangle^{(a)} \otimes U_{x,\pm}^{(a)} |\tau_{b_0}\left(\tau_a\circ\tau_{b_0} (G)-a\right)\rangle\, ,
\end{eqnarray}
where the local unitaries $U_{i,\pm}^{(a)}$ are defined as for Proposition~\ref{pauli}.  

\section{Proofs}\label{Proofs}

In this section we prove the statements that for
clarity of presentation have been summarized
in the previous section without proof. 

{\it Proof of Proposition \ref{pauli}:}
As already mentioned in Sec.~\ref{LU classes}, with the LU-rule at hand one could derive
the graph $G'$ after an $x$- or $y$-measurement from the $z$-measurement rule,
which can be directly proven by disentangling the Ising interactions $U^{(a,b)}$ in 
Eq.~(\ref{Ising}). Here we will instead take another 
starting point for the proof, namely a well-known
result from stabilizer theory that
we will then apply to
the specific question at hand.
Consider a subspace of ${\mathbbm{C}}^n$
which is stabilized by
\begin{equation}
\langle \{g_i\}_{i\in I} \rangle,  \,\,\,\,\,\,\,\,\,\,\,\,
g_i\in \mathcal{P}_n,
\end{equation}
where $\mathcal{P}_n$ denotes the Pauli group
on $n$ qubits and $I$ is an index set.
It is well known (see, e.g.,  Ref.~\cite{NielsenBook})
that the projected subspace
$P_{g,\pm}$ corresponding to a measurement of an operator
$g\in {\cal P}_n$ in the Pauli group (i.e.,  $g$ is a product of
Pauli matrices)
 with outcome $\pm 1$
is stabilized by
\begin{itemize}
\item[(i)] $\langle \{g_i\}_{i \in I} \rangle$,
if $g$ commutes with
all stabilizer generators $g_i$.
\item[(ii)] $\langle \lbrace \pm g\rbrace \cup \lbrace
{g_k g_j \, : \, j \in I' - \{k\}}
\rbrace \cup \lbrace g_j \, | \,
j \in I'^c \rbrace \rangle$ for some $k \in I'$ otherwise.
$I'$ denotes the non-empty index set of the generators
$g_i$ that do not commute with $g$, and $I'^{c}=I\backslash
I'$ is
the complement of $I'$.
\end{itemize}

We now turn to the specific case of graph state vectors
 $| G \rangle$ and measurements of $\sigma_x^{(a)},
\sigma_y^{(a)}$ or $\sigma_z^{(a)}$ at vertex $a \in V$.
Then each generator $K_G^{(a)}$ is associated with an element
$a \in V$, and for a given $g$, the list $I'$ of generators that do
not commute with $g$ is a subset of $V$. For the measurements
considered here, only case (ii) is relevant,
 as long as $\sigma_x^{(a)}$ is not measured at an isolated vertex
$a$.
 In the latter situation, which corresponds to case (i),
\begin{equation}
 K_G^{(a)}=\sigma_x^{(a)},
\end{equation}
and $\sigma_z^{(a)}$
is not contained in any $K_G^{(b)}$ for $b\not= a$. Then
the state is left unchanged and with probability $1$ the result
$+1$ is obtained.

In case (ii), in turn, the possible measurement results $\pm 1$ are
 always obtained each with probability $1/2$.
Let us start with identifying the resulting state
vector and graph after
measuring $\sigma_{z}^{(a)}$. The index set $I'$
then is given by $I'=\{a\}$, and
the state vector
$P^{(a)}_{z,\pm}|G\rangle$ is stabilized by
\begin{equation}
\langle \lbrace \pm \sigma_z^{(a)}\rbrace  \cup \lbrace K_G^{(b)} \,
: \, b \in V - \{a\} \rbrace \rangle.
\end{equation}
Multiplying
$\pm \sigma_z^{(a)}$ to the elements $K_G^{(b)}$
for $b\in V - \{a\} $, according to the neighbors $b \in N_a$ in $G$, yields
\begin{equation}
\pm \sigma_z^{(a)}K_G^{(b)}=\pm \sigma_x^{(b)} \prod_{b' \in
N_b - \{a\}}\sigma_z^{(b')},
\end{equation}
which is up to the sign the stabilizer
generator according to the vertex $b$ in $G-\{a\}$.
Since the stabilizer generators $K_G^{(b)}$ corresponding
to vertices
$b$ outside $N_a \cup \{a\}$ in $G$ coincide with those in
$G-\{a\}$, the
stabilizer may as well be seen generated by
\begin{eqnarray}
\lbrace \pm \sigma_z^{(a)}\rbrace  & \cup &
\lbrace \pm K_{G-\{a\}}^{(b)}
\, : \, b \in N_a \rbrace \nonumber\\
&\cup&
\lbrace K_{G-\{a\}}^{(b)} \, : \, b \in
V-\{a\}-N_a \rbrace \, .
\end{eqnarray}
Hence, we have shown
the validity of Eq.~(\ref{ProjectiveM})
for the case of a
positive $\sigma_z$ measurement result.
In the other case the sign can be corrected for,
as the stabilizer can be written as
\begin{eqnarray}
    \langle \, U_{z,-} \left( \lbrace - \sigma_z^{(a)}\rbrace  \cup
\lbrace K_{G-\{a\}}^{(b)} \, : \, b \in V_{G-\{a\}} \rbrace \right)
U_{z,-}^\dagger \, \rangle \, ,
\end{eqnarray}
which corresponds to the state vector
\begin{equation}
|z,-\rangle^{(a)} \otimes U_{z,-}^{(a)}
|G-\{a\}\rangle.
\end{equation}
Here, it has been used that
$
    U_{z,-}^{(a)} = \prod_{b \in N_a} \sigma_z^{(b)}$
anti-commutes
exactly with the generators
\begin{equation}\lbrace K_{G-\{a\} }^{(b)} \, : \, b \in N_a
\rbrace .
\end{equation}

In a similar manner, the case of a measurement of
$\sigma_{y}^{(a)}$ can be treated. The index set $I'$
is given by
$I'=N_a\cup\{a\}$ and, if $k =a$ is chosen,
the new stabilizer is given by
\begin{eqnarray}
\langle \,\lbrace \pm \sigma_y^{(a)}\rbrace
\cup {\mathcal G}_1 \cup
{\mathcal G}_2\, \rangle,
\end{eqnarray}
where
\begin{eqnarray}
{\mathcal G}_1 & =& \lbrace K_{G}^{(a)} K_{G}^{(b)} \, : \, b
\in N_a
\rbrace,\\
{\mathcal G}_2 & =&\lbrace K_{G}^{(c)} \, : \, c \in V-  N_a
-\{a\} \rbrace .
\end{eqnarray}
For ${\mathcal G}_1$ one computes
\begin{eqnarray}
 &   & K_{G}^{(a)} K_{G}^{(b)}\, = \, \sigma_y^{(a)}\sigma_y^{(b)}
\prod_{b' \in N_b\Delta N_a-\{a,b\}} \sigma_z^{(b')} \cr
 & = & \pm \sigma_y^{(a)}\, U_{y,\pm}^{(a)} \left( \sigma_x^{(b)}
 \prod_{b'
\in N_b\Delta N_a-\{a,b\}} \sigma_z^{(b')} \right)
{U_{y,\pm}^{(a)}}^\dagger
\cr
 & = & \pm \sigma_y^{(a)} U_{y,\pm}^{(a)} K_{G'}^{(b)} {U_{y,\pm}^{(a)}}^\dagger ,
\end{eqnarray}
where $G'$ denotes the graph with the edge set
$E'=E_G\Delta
E(N_a,N_a)$ and the unitaries
$U_{y,\pm}^{(a)}$ are defined as in Eq.~(\ref{uy}).
Because the elements in ${\mathcal G}_2$ commute with $U_{y,\pm}^{(a)}$, we
arrive at the result for measurements of $\sigma_{y}^{(a)}$.

Finally, in the case of measurements of $\sigma_{x}^{(a)}$, we
identify $I'$ as
$I'=N_a$. If some $b_0 \in N_a$ is chosen, the new stabilizer is
given by
\begin{eqnarray}
\langle \,\lbrace \pm \sigma_x^{(a)}, K_{G}^{(a)} \rbrace \cup
{\mathcal G}_1 \cup {\mathcal G}_2 \cup {\mathcal G}_3 \cup {\mathcal
G}_4  \, \rangle,
\end{eqnarray}
where, because of the
following argumentation the finer dissection is chosen,
\begin{eqnarray}
{\mathcal G}_1 & = & \lbrace K_{G}^{(b_0)} K_{G}^{(b)} \, : \, b \in
N_a\cap N_{b_0} \rbrace ,\\
{\mathcal G}_2 & = & \lbrace K_{G}^{(b_0)} K_{G}^{(b)} \, : \, b \in
N_a-  N_{b_0}- \{b_0\}  \rbrace ,\\
{\mathcal G}_3 & = & \lbrace K_{G}^{(b)} \, : \, b \in N_{b_0}-
 N_a - \{a\}  \rbrace ,\\
{\mathcal G}_4 & = & \lbrace K_{G}^{(c)} \, : \, c \in V -
N_a - N_{b_0}  \rbrace \, .
\end{eqnarray}
Instead of $K_{G}^{(a)}$, the generator
\begin{eqnarray}
    \pm
\sigma_x^{(a)}K_{G}^{(a)} &=& \pm \;\prod_{b \in N_a} \sigma_z^{(b)}
\nonumber\\
&=&
U_{x,\pm}^{(a)}\,\left(\sigma_x^{(b_0)}\prod_{b \in N_a-\{b_0\}}
\sigma_z^{(b)} \right)\, {U_{x,\pm}^{(a)}}^\dagger
\end{eqnarray}
can be chosen,
where $U_{x,\pm}^{(a)}$ is defined as above.
Instead of $K_{G}^{(b_0)} K_{G}^{(b)}$
in ${\mathcal G}_1$, we choose, for
$b_0 \in N_a$ and $b \in N_{b_0}$,
\begin{eqnarray}
 &&  \pm\, \sigma_x^{(a)}K_{G}^{(a)} K_{G}^{(b_0)}
 K_{G}^{(b)}
  = \mp \, \sigma_x^{(b_0)}
\sigma_x^{(b)}  \prod_{b' \in N_b \Delta N_a \Delta N_{b_0}}
\sigma_z^{(b')} \cr
 & = &  U_{x,\pm}^{(a)} \biggl( \;\mp \,(\mp i \sigma_y^{(b_0)})
\sigma_x^{(b_0)} (+\sigma_x^{(b)})
\prod_{b' \in N_b \Delta N_a \Delta N_{b_0}  }
\sigma_z^{(b')}\;\biggr) {U_{x,\pm}^{(a)}}^\dagger \cr
 & = &  U_{x,\pm}^{(a)}
 \left( \sigma_x^{(b)} \prod_{b' \in N_b \Delta N_a
\Delta N_{b_0} \cup \{b_0\} } \sigma_z^{(b')}\right)
{U_{x,\pm}^{(a)}}^\dagger,\nonumber
\end{eqnarray}
where the second equality holds, because $b_0 \not\in N_b \Delta N_a
\Delta N_{b_0}$ and $(\mp i \sigma_x^{(b_0)})^{1/2}$
therefore anti-commutes only with $\sigma_x^{(b_0)}$.
Moreover, the positive sign of $+\sigma_x^{(b)}$ is due to $b \not\in
N_a -N_{b_0} -\{b_0\} $, as well as $b \not\in N_{b_0} - N_a - \{a\} $,
since in both cases $\pm$ the term $\sigma_z^{(b)}$ of $U_{x,\pm}^{(a)}$
commutes with $\sigma_x^{(b)}$.
For $K_{G}^{(b_0)} K_{G}^{(b)}$ of ${\mathcal G}_2$ one computes,
for $b\not\in N_{b_{0}}$, $b_{0} \not\in N_{b}\Delta N_{b_{0}}$,
$b\in N_{a}-N_{b_{0}}-\{b_0\}$,
\begin{eqnarray}\nonumber
 &   &  K_{G}^{(b_0)} K_{G}^{(b)}  \cr
 & = & \sigma_x^{(b_0)} \sigma_x^{(b)}
\prod_{b' \in N_b\Delta N_{b_0}} \sigma_z^{(b')} \cr
 & = &  U_{x,\pm}^{(a)}
\biggl(\; (\mp i \sigma_y^{(b_0)}) \sigma_x^{(b_0)} (\mp
\sigma_x^{(b)})\; \prod_{b' \in N_b  \Delta N_{b_0}  }
\sigma_z^{(b')}\;
\biggr)\, {U_{x,\pm}^{(a)}}^\dagger \\
 & = &  U_{x,\pm}^{(a)} \left( \sigma_x^{(b)} \prod_{b' \in N_b \Delta
N_{b_0} \cup \{b_0\} } \sigma_z^{(b')}\right) {U_{x,\pm}^{(a)}}^\dagger  \, .
\end{eqnarray}
Instead of $ K_{G}^{(b)}$ in ${\mathcal G}_3$ we choose,
for $b \not\in N_{a}$, $b_0 \not\in N_{b}\Delta N_a $,
$b \in
N_{b_0}-N_a-\{a\}$,
\begin{eqnarray}\nonumber
 \pm \sigma_x^{(a)} K_{G}^{(a)} K_{G}^{(b)}
 &  = & \pm \,\sigma_x^{(b)}  \prod_{b' \in N_b\Delta N_{a}}
\sigma_z^{(b')} \cr
 &  = &
U_{x,\pm}^{(a)}\left(
\pm \,(\pm \sigma_x^{(b)}) \prod_{b' \in N_a  \Delta N_{b}  }
\sigma_z^{(b')}\right) {U_{x,\pm}^{(a)}}^\dagger \\
 & = &  U_{x,\pm}^{(a)} \left( \sigma_x^{(b)}
 \prod_{b' \in N_b \Delta N_{a}
} \sigma_z^{(b')}\right) {U_{x,\pm}^{(a)}}^\dagger  \, .
\end{eqnarray}
Moreover, note that $ K_{G}^{(c)}$ in ${\mathcal G}_4$ is not changed
by $U_{x,\pm}^{(a)}$, since $c \in V - \left( N_a \cup N_{b_0} \right)$.
To summarize, the new neighborhoods $N'_b$ are
\begin{eqnarray}
N'_b = \left \{ \begin{array}{ccc}
N_a - \{b_0\} & \text{if}&  b=b_0 ,\\
N_b\Delta N_a \Delta N_{b_0} \cup\{b_0\} & \text{if} & b\in N_{b_0}
\cap N_a, \\
N_b \Delta N_{b_0} \cup \{b_0\} & \text{if} & b\in N_{a}
- N_{b_0} - \{b_0\}  ,\\
N_b \Delta N_{a}  &\text{if} & b\in N_{b_0} - N_{a} -
\{a\} , \\
N_b &\text{if} &  b\in V_{G} - N_{a} - N_{b_0}.\\
\end{array} \right .
\end{eqnarray}
A comparison shows that these neighborhoods
correspond exactly to the graph $G'$
obtained from Eq.~(\ref{graph x measurement}).
This concludes the proof.
\proofend

{\it Proof of Proposition \ref{perst}:} This statement
follows immediately
from Eq.~(\ref{Persistency}) in property (ii) of the Schmidt measure, and the fact that the
different measurement results are obtained with probability $1/2$.
\proofend

{\it Proof of Proposition \ref{bip}:}
To show Eq.~(\ref{partial trace}), the partial trace over
$A$ can be taken according to the basis of $A$
given by
\begin{equation}
\left\{|{\mathbf z} \rangle=
\bigotimes_{a \in A} | z ,
(-1)^{z_a}\rangle^{(a)}\right\}.
\end{equation}
This corresponds to successive local $\sigma_z$-measurements of all
vertices in $A$, yielding measurement outcomes
$\pm 1$.
According to Sec.~\ref{Pauli measurements},
after measurement of  $\sigma_z^{(a)}$
the state of the remaining vertices is the graph state vector
$| G - \{a\} \rangle$ in the case
of the outcome $+1$ , and
\begin{equation}\prod_{c\in N_a} \sigma_z^{(c)}
| G -\{a\}  \rangle,
\end{equation}
if the outcome is $-1$.
This can be summarized as
\begin{equation}
    \left(\prod_{c\in N_a}
    \sigma_z^{(c)}\right)^{z_a} | G -\{a\}
    \rangle,
\end{equation}
where $z_a \in \{0,1\}$
denotes the measurement result $\pm 1$.
Since the following measurements commute with the previous local
unitaries,
the final state vector
according to the result ${\mathbf z}=(z_a)_{a\in A}\in
{\mathbbm F}_2^A$ is
\begin{eqnarray}\nonumber
&   & \prod_{a \in A}\prod_{c \in N_a} \left( \sigma_z^{(c)}
\right)^{z_a}|{\mathbf z} \rangle\otimes | G - A\rangle\cr
& = & \prod_{a \in A}\prod_{c \in V}
\left(\sigma_z^{(c)}\right)^{\Gamma_{ca}z_a} |{\mathbf
z}\rangle\otimes | G - A\rangle\cr
& = & \prod_{a \in A} \left(\sigma_z^{(a)}\right)^{\langle {\mathbf
e}^{a}| \Gamma_{G-B} {\mathbf z}\rangle} |{\mathbf z}\rangle \,
\otimes\\
&   & \prod_{b \in B} \left(\sigma_z^{(b)}\right)^{\langle {\mathbf
e}^{b}| \Gamma_{AB} {\mathbf z} \rangle} | G - A\rangle ,
\end{eqnarray}
where  the computation with respect to ${\mathbf z}$ is done in
${\mathbbm F}_2^A$ (i.e.,
modulo 2) and ${\mathbf e}^{b}_a =
\delta_{ab} $. Therefore, we arrive at the resulting
state state vector associated with the result
${\mathbf z}$ as
\begin{eqnarray}
(-1)^{\langle{\mathbf z} | \Gamma_{G-B } {\mathbf z} \rangle}
\, |{\mathbf z}\rangle  \otimes
\prod_{b \in B} \left(\sigma_z^{(b)}\right)^{\langle {\mathbf
e}^{b}| \Gamma_{AB} {\mathbf z} \rangle} | G - A\rangle .
\end{eqnarray}
Because the possible measurement results are attained with
probability $1/2$,
this proves the validity of Eq.~(\ref{partial trace})
with  local unitaries as in Eq.~(\ref{U in partial trace}),
i.e.,
\begin{eqnarray}
U({\mathbf z})= \prod\limits_{a \in A} \left( \prod \limits_{b\in
N_a} \sigma_z^{(b)} \right)^{z_a} = \prod_{b \in B}
\left(\sigma_z^{(b)}\right)^{\langle {\mathbf e}^{b}| \Gamma_{AB}
{\mathbf z} \rangle}  \, .
\end{eqnarray}

To show the validity of Eq.~(\ref{Schmidt rank}),
note that for any
${\mathbf z}_1,{\mathbf z}_2 \in {\mathbbm F}_2^A$,
the state vectors
$U({\mathbf z}_1)| G - A\rangle$ and $U({\mathbf z}_2)| G - A\rangle$
are orthogonal if and only if
\begin{eqnarray}
U({\mathbf z}_1 - {\mathbf z}_2) = U({\mathbf
z}_2)^\dagger U({\mathbf z}_1) \not= {\mathbbm{1}},
\end{eqnarray}
since   $\prod_{c \in
V'} \sigma_z^{(c)}$ anti-commutes with the stabilizer for any graph
state and for any $\emptyset \not= V'\subseteq V $,
and  therefore takes it into its orthogonal complement. Hence,
 \begin{eqnarray}\nonumber
&   & {\rm log}_2 \left( \text{rank} \left( \text{tr}_A [| G \rangle
\langle G | ]\right) \right) \nonumber \\
& = &  {\rm log}_2  \left(\text{dim} \, \text{span} \,\lbrace
U({\mathbf z})| G - A\rangle  : {\mathbf z} \in {\mathbbm F}_2^A
\rbrace \right) ,
\end{eqnarray}
as for every ${\mathbf z} \in {\mathbbm F}_2^A
$ exactly those  ${\mathbf z'} \in {\mathbbm F}_2^A $  yield the
same
$U({\mathbf z'})=U({\mathbf z})$, for which
\begin{equation}
\mathbf z' - \mathbf z
\in \lbrace {\mathbf z} \in {\mathbbm F}_2^A \, :  \, U({\mathbf z})=
{\mathbbm{ 1}} \rbrace
\end{equation}
holds.
This gives
\begin{eqnarray}
&&{\rm log}_2 \left( \text{rank} \left( \text{tr}_A [| G \rangle
\langle G | ]\right) \right)
\nonumber\\
& = & |A| - {\rm log}_2 |\lbrace {\mathbf z} \in {\mathbbm F}_2^A \,:
\, \langle {\mathbf e}^{b}| \Gamma_{AB} {\mathbf z} \rangle
=_{{\mathbbm F}_2} 0 \; \forall b\in B \rbrace| \nonumber \\
& = & |A| - \text{dim} \,\text{ker}_{{\mathbbm F}_2}(\Gamma_{AB} ) =
\text{rank}_{{\mathbbm F}_2} (\Gamma_{AB} ).
\end{eqnarray}

\begin{figure}[th]

\includegraphics[width=5cm]{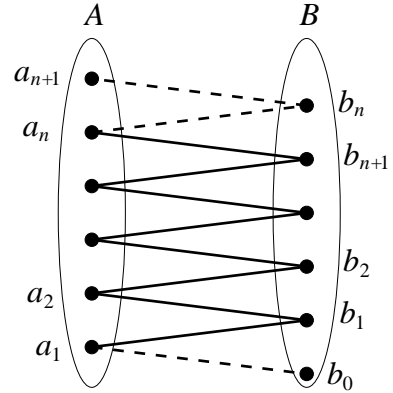}
\caption{\label{fig:MaxSchmidtRank} A sufficient 
condition for a graph to have maximal Schmidt rank.}
\end{figure}
{\it Proof of Proposition \ref{sufficient crit for max rank}:}
To see this, assume to the contrary that $G_{AB}$
contains no cycles but that the Schmidt rank is not maximal. Then, denote with $A'\subseteq A$
any subset for
which the corresponding columns $\mathbf{n}^{(a)}$ in $\Gamma_{AB}$ might
add to $0$ modulo 2,
\begin{equation}\label{dependent neighborhoods}
\sum_{a \in A'} \mathbf{n}^{(a)} =_{{\mathbbm F}_2} 0 \; .
\end{equation} Obviously,
every vertex $b \in B' = \bigcup_{a \in A'}N_a$
must have an even number of distinct neighbours in $A'$. For the moment let the single leaf $a_1$ be contained in $A'$  and
\begin{equation}
    a_1, b_1,a_2,...,b_{n-1},a_n,
\end{equation}
be a $\{a_{1},a_{n}\}$-path with maximal length that
alternately crosses the sets $A'$ and
$B'$ (starting in $a_1$ and ending in $A'$ as depicted in Fig. \ref{fig:MaxSchmidtRank}).
Because $a_n$ is necessarily a vertex of degree more than $1$ in $G_{AB}$ and by construction also in
$G_{A'B'}$, it must have a neighbor $b_n\not= b_{n-1}$ in $B'$. If
$b_n=b_i$ for some $i=1,...,n-2$,  a contradiction is found. Otherwise
$b_n$ itself must have a neighbor $a_{n+1}\not= a_n$ in $A'$,
because $b_n$ has even degree in $G_{A'B'}$. Now either
$a_{n+1}=a_{i}$ for some $i=1,...,n-2$ or the path
\begin{equation}
    a_1,b_1,a_2,...,b_{n-1},a_n,b_n,a_{n+1}
\end{equation}
is a longer path in $G_{A'B'}$, both yielding to contradictions with
the previous assumptions.

 If the single leaf $a_1$ is not contained in $A'$, or if $A$ contains no
 leaves, the previous argumentation still holds, because now any $a\in A'$
 must have a degree more than one, if one allows $a_1\in A'$ to be arbitrary.
 The sufficient criterion for the connected components of $G_{AB}$ then
 follows from the additivity of $E_S$ within the given bipartition $(A,B)$, as
 formulated in Eq.~(\ref{E additivity for bipartitions}), after
 deleting all edges within $G[A]$ and $G[B]$, which is proper
 $(A,B)$-local unitary operation.
\proofend

\begin{figure}[th]

\includegraphics[width=6cm]{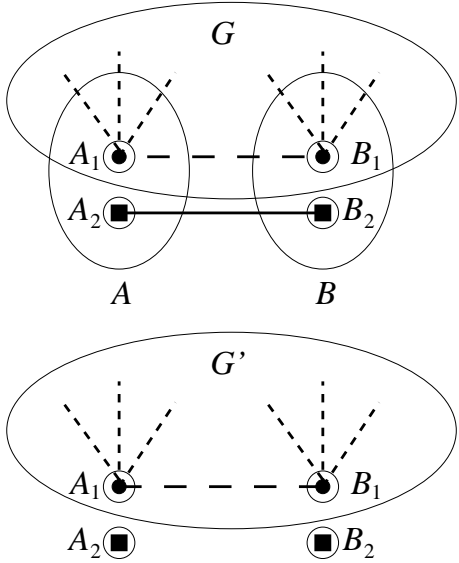}
\caption{\label{fig:AddEdge} The situation before and after the LOCC simulation for
adding or deleting an edge $\{a_1,b_1\}$: the graph state vector
$|G\rangle$ can be transformed
by $(A,B)$-local operations and classical communication with probability $1$ into the
state vector $|G'\rangle$, where the edge between the partitions $A_1$ and $B_1$ is added or deleted. This
is possible
if one allows for an additional maximally entangled state
$\vrule height6pt width6pt depth2pt \vrule height1pt width20pt depth0pt \vrule height6pt
width6pt depth2pt$ between $A_2$ and $B_2$. After the LOCC operation the resource is consumed, i.e.,
the state of $(A_2,B_2)$ is a pure product state
$\vrule height6pt width6pt depth2pt \vrule height0pt width20pt depth0pt \vrule height6pt width6pt depth2pt\,$.}
\end{figure}

{\it Proof of Proposition \ref{edger}:} Let $G=(V\cup\{ a_1, b_1\}
,
E)$ be a graph. The set $V$ is the set of all vertices of the graph
$G$ except the two
vertices $a$ and $b$
between which an edge is supposed to be deleted or
added. Let $V$ also denote the sequence of partitions in the finest partitioning of $G$ and $A_1=\{a_1\}$, $B_1=\{b_1\}$.
 $G'$ denotes the resulting graph, which differs from $G$ in the
edge $\{a_1,b_1\}$. As has been shown in Refs.
\cite{Simulate0,Simulate,Simulate1}, the
unitary operation
corresponding to the Ising interaction, see Eq.~(\ref{IsingInteraction}),
can be implemented with LOCC with unit probability. The necessary and
sufficient resources are one maximally entangled pair of
qubits and one bit of classical communication in each direction (see Fig. \ref{fig:AddEdge}).
The vertices $a_2$ and $b_2$ correspond to the qubits that
carry the entanglement $|\psi\rangle$ resource required to implement the Ising
interaction with LOCC. With $A_2=\{a_2\}$ and  $B_2=\{b_2\}$ we can conclude that
\begin{eqnarray}
    &&E_{S}^{(V,A_1, B_1)}(|G\rangle) +1 \nonumber\\
    &=&
    E_{S}^{(V,A_1, B_1)} (|G\rangle) +
    E_{S}^{(A_2,B_2)}(|\psi\rangle)\nonumber\\
    &\geq & E_{S}^{(V,A_1, B_1, A_2, B_2)}(|G\rangle
\otimes
    |\psi\rangle),
\end{eqnarray}
due to sub-additivity and
\begin{eqnarray}
 E_{S}^{(V,A_1, B_1, A_2,B_2)}(|G\rangle \otimes
    |\psi\rangle)
    &\geq & E_{S}^{(V,A, B )}(|G\rangle \otimes
    |\psi\rangle),\nonumber\\
\end{eqnarray}
due to
 the non-increasing property under
coarse graining of the partition $ A=A_1\cup A_2$ and $ B=B_1\cup B_2$. As the Schmidt measure is
an entanglement monotone,
LOCC simulation of the Ising interaction yields
\begin{eqnarray}
    & & E_{S}^{(V,A_1, B_1)}(|G\rangle) +1 \nonumber\\
    &\geq & E_{S}^{(V,A,B)}(|G'\rangle\otimes
    |\phi\rangle^{(a_2)}\otimes
|\omega\rangle^{(b_2)})\nonumber\\
    &= & E_{S}^{(V,A_1,B_1)}(|G'\rangle),
\end{eqnarray}
where it has been used that local additional systems can always
be appended without change in the Schmidt measure. The state vector
$|\phi\rangle^{(a_2)}\otimes |\omega\rangle^{(b_2)}$
corresponds to the state vector of the additional system after
implementation of the Ising gate. Since the Ising
interaction gives rise to both a deletion or the addition of an edge,
we have arrived at the above statement.
Note that the whole argumentation also holds
if $a_1$ and $b_1$ are
vertices in some coarser partitions $A_1$ and $B_1$ of $G$. In this
case the same simulation with LOCC of the Ising interaction can be
used, but in the estimations now with respect to coarser partitions.
\proofend

{\it Proof of Proposition \ref{vertexr}:}
If a vertex $a\in V$
is deleted from a graph $G=(V,E)$,
the corresponding graph state vector  $| G-\{a\}
\rangle$ is according to Proposition \ref{pauli} up to local
unitaries the graph state that is obtained from a measurement of
$\sigma_z^{(a)}$ at the vertex $a$. According to Eq.~(\ref{E under
SLOCC})
the Schmidt measure cannot increase, and because of Eq.~(\ref{E under projective measurement}) it can at most decrease by
one.
\proofend

{\it Proof of Proposition \ref{twoc}:}
To see this, we can write the adjacency matrix
$\Gamma_G$ according to the partitions of sources $A$ and sinks
$B$. Then, for $\Gamma_G$ in Eq.~(\ref{Gamma for bipartition})
\begin{equation}
\Gamma_{G[A]} \,=\,\Gamma_{G[B]}\, = \, 0,
\end{equation}
and the number of
linearly independent columns/rows in $\Gamma_G$ is twice that of
$\Gamma_{AB}$. Hence, a lower bound is
\begin{equation}
E_S^{(A,B)}(|G\rangle) =
\lfloor \frac{1}{2} \text{rank}_{{\mathbbm F}_2}(\Gamma_{G})\rfloor.
\end{equation}
If $\Gamma_G$ is invertible, then
\begin{equation}
E_S(|G\rangle)\geq\lfloor \frac{|V|}{2} \rfloor
\end{equation}
holds.
On the other hand, each of the partition $A$ and $B$ is a vertex cover
of $G$ and $E_S(|G\rangle)$ is therefore bound from above by the
size of the smaller partition, which must be less than
$\lfloor |V| / 2 \rfloor$.
\proofend

{\it Proof of Proposition \ref{loc}:}
Let $c \in V-N_a$, then
\begin{equation}
    UK_G^{(c)}U^\dagger=K_G^{(c)}=K_{G'}^{(c)}.
\end{equation}
For $b \in N_a$, one computes
\begin{eqnarray}\nonumber
 &   &  U\,K_G^{(b)}\,U^\dagger   \cr
 & = &  \left( i \sigma_z^{(b)}\right) \,\sigma_x^{(b)} \; \left(- i
\sigma_x^{(a)}\right) \,\sigma_z^{(a)} \prod_{b' \in N_b - \{a\}}
\sigma_z^{(b')} \cr
 & = & \sigma_x^{(a)}\,\prod_{b' \in N_a} \sigma_z^{(b')}\; \cdot \;
\sigma_x^{(b)}\,\prod_{b'' \in N_b \Delta N_a} \sigma_z^{(b'')} \\
 & = &  K_{G'}^{(a)}\; \cdot \; K_{G'}^{(b)}  \, .
\end{eqnarray}
Therefore,
\begin{eqnarray}
    \langle UK_G^{(c)}U^\dagger\rangle_{c\in V} = \langle
    K_{G'}^{(c)}\rangle_{c\in V},
\end{eqnarray}
which had to be shown.

\proofend

\section{Examples}\label{Examples}

In this section the findings of the previous two sections will be
applied to evaluating the Schmidt measure for a number of important
graph states. Upper and lower bounds will be investigated, and in most
of the subsequently considered cases, these bounds coincide, hence
making the computation of this multi-particle entanglement measure
possible.

\begin{widetext}

\begin{figure}[th]
\includegraphics[width=12cm]{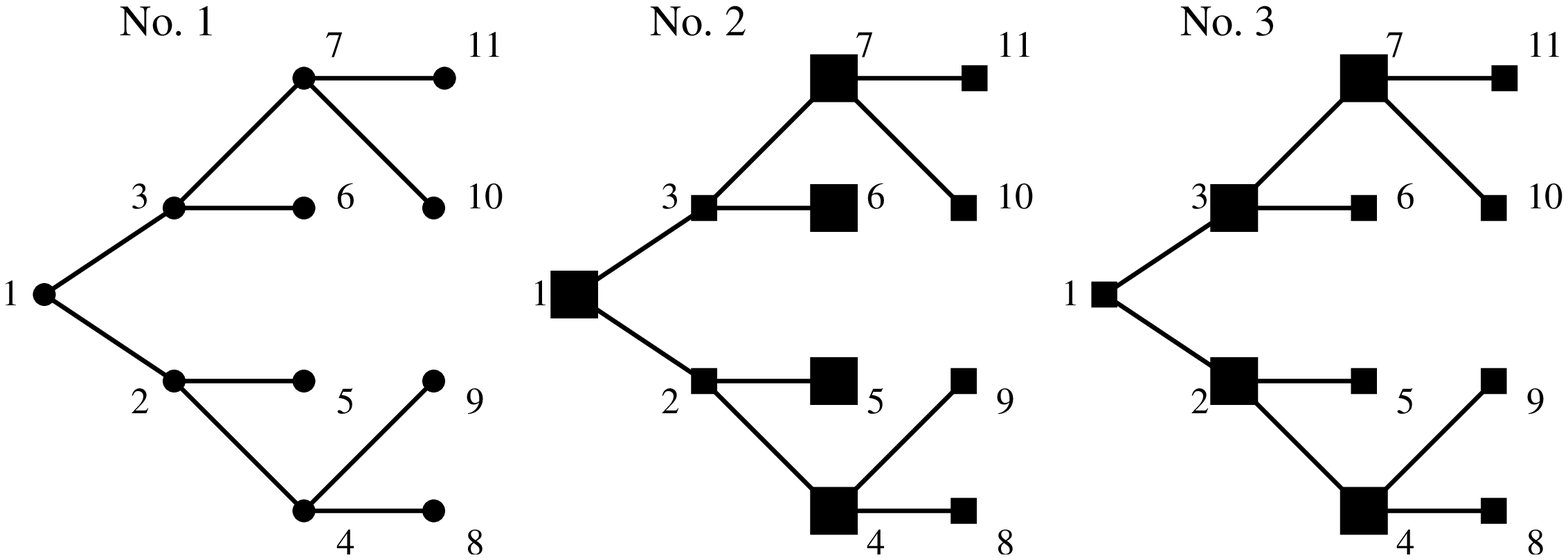}
\caption{\label{fig:TreeExample1} The graph No.~1 
represents a tree. Its bipartitioning $(A,B)$, for which  in graph
No.~2 the vertices in $A$ are depicted by large boxes 
$\vrule height7pt width7pt depth0pt$, is neither a
minimal vertex cover nor yields maximal partial rank.
Instead the set of vertices $A$, represented by large boxes
$\vrule height7pt width7pt depth0pt$ in graph No.~3, is a 
minimal vertex cover with maximal partial rank.
Here, the edges within in the set $A$ are drawn by thin lines in order to
illustrate the resulting graph $G_{AA^c}$ between $A$ and
its complement, as considered in Sec.~\ref{E_S for bipartite splits}. }
\end{figure}

\end{widetext}

{\it Example 1:
The Schmidt measure of a tree is the size of its smallest
vertex cover.}

{\it Proof:} A {\it tree} is a graph that has no cycles.
We claim that a minimal vertex
cover $A$ of $G$ can be chosen, such that the graph $G_{AB}$
between $A$ and
its complement $B=A^c$ fulfils the sufficient criterion in Proposition \ref{sufficient crit for
max rank} for maximal Schmidt rank.
To see this, let $A$ be a minimal vertex cover.
If a connected component $C_1$ of $G_{AB}$ has more than one leaf $a$ in $A\cap C_1$,
then this can be transferred to another (possibly new) component $C_2$, by simply
exchanging the leaves in $A$
with their unique neighbors $b$
in $B$. One again obtains a vertex cover of the same (hence
minimal)
size. Note that by this exchange the new complement $B'$ receives
no inner edges with respect to $G$,
since each of the exchanged vertex of $A$ only had one neighbor
in $B$.

Two distinct leaves $a_2$ and $a_3$ in $A$ cannot be adjacent to the same vertex $b\in B$. Otherwise, taking $b$ instead of both $a_2$ and $a_3$ in $A$ would  yield a vertex cover with fewer vertices.
Moreover, two distinct leaves $a_2$ and $a_3$ of $A\cap C_1$  are necessarily
transferred to different connected components $C_2$ and $C_3$ of $G_{AB}$, because otherwise
any two elements $a_2'$ and $a_3'$ of $N_{a_2}\cap A$ and $N_{a_3}\cap A$ are
connected by an $(A,B)$-path, which together with an $(A,B)$-path between $a_2$
and $a_3$ and the edges $\{a_2,a_2'\}$ and $\{a_3,a_3'\}$ would form a cycle of $G$.

Starting with a component $C'_1$ apart from one leaf $a_1$, all other leaves
$a_2,...,a_k$ can be transferred in this way to different components $C'_2,...,C'_k$.
Let us fix these vertices, including their unique neighbors $b_1,...,b_k$, for the following
reduction of the number of leaves in the components $C'_2,...,C'_k$ in the sense
that only vertices which differ from $a_1,...,a_k,b_1,...,b_k$, are considered for
a subsequent transfer. Since $G$ is free of cycles, similar to the above argument, none
of the remaining leaves is transferred to a component which was already obtained by
previous transfer.
 In a similar manner, for all remaining components $C$ the minimal vertex cover
 can be transformed into a new one $A'$, for which $C\cap A'$ contains only one
 leaf without affecting components which were already considered in the transfer
  process. That shows the validity of our claim.
\proofend

\begin{widetext}

\begin{figure}[th]
\includegraphics[width=8.5cm]{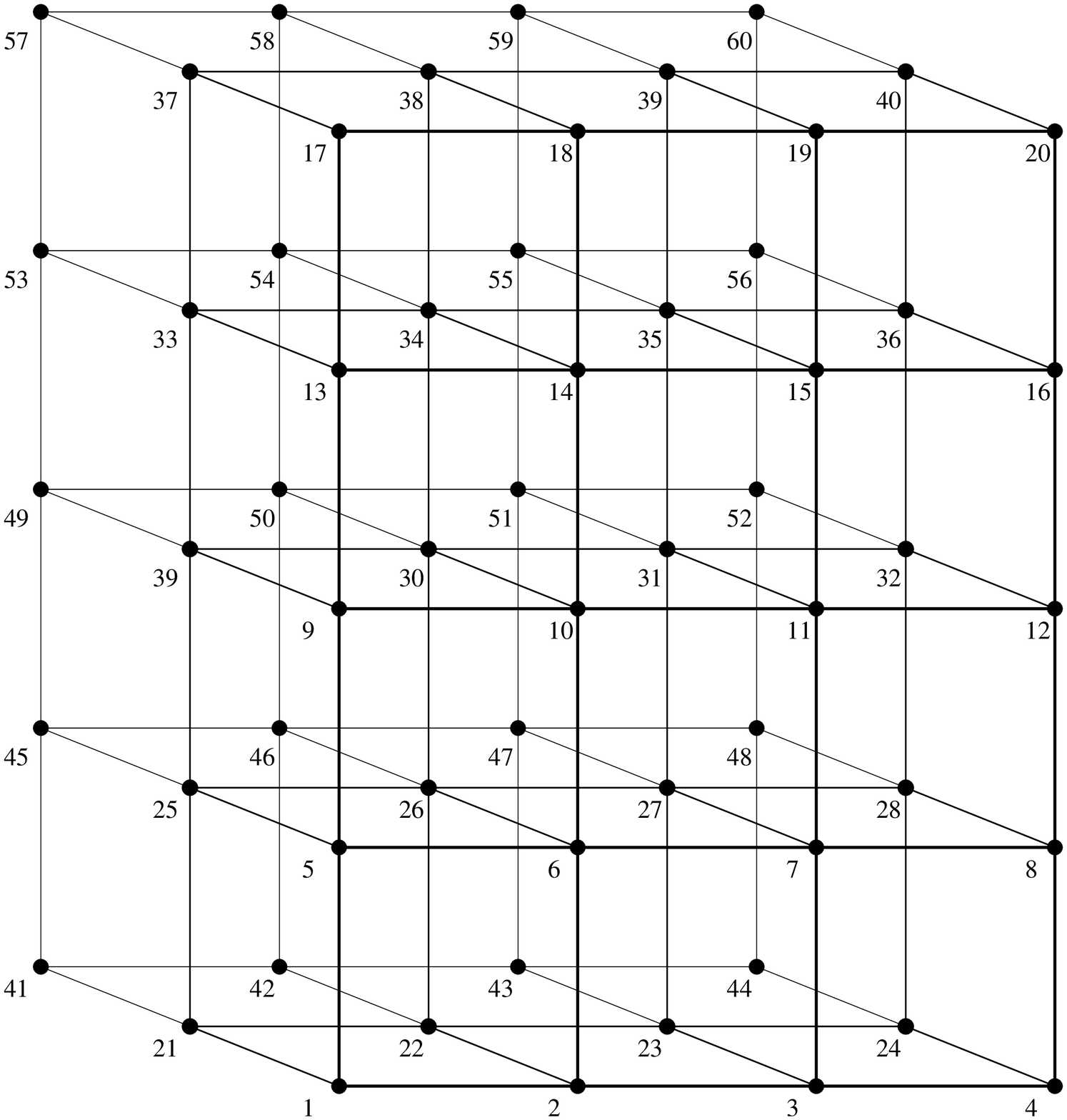}\hspace*{.5cm}\includegraphics[width=8.5cm]{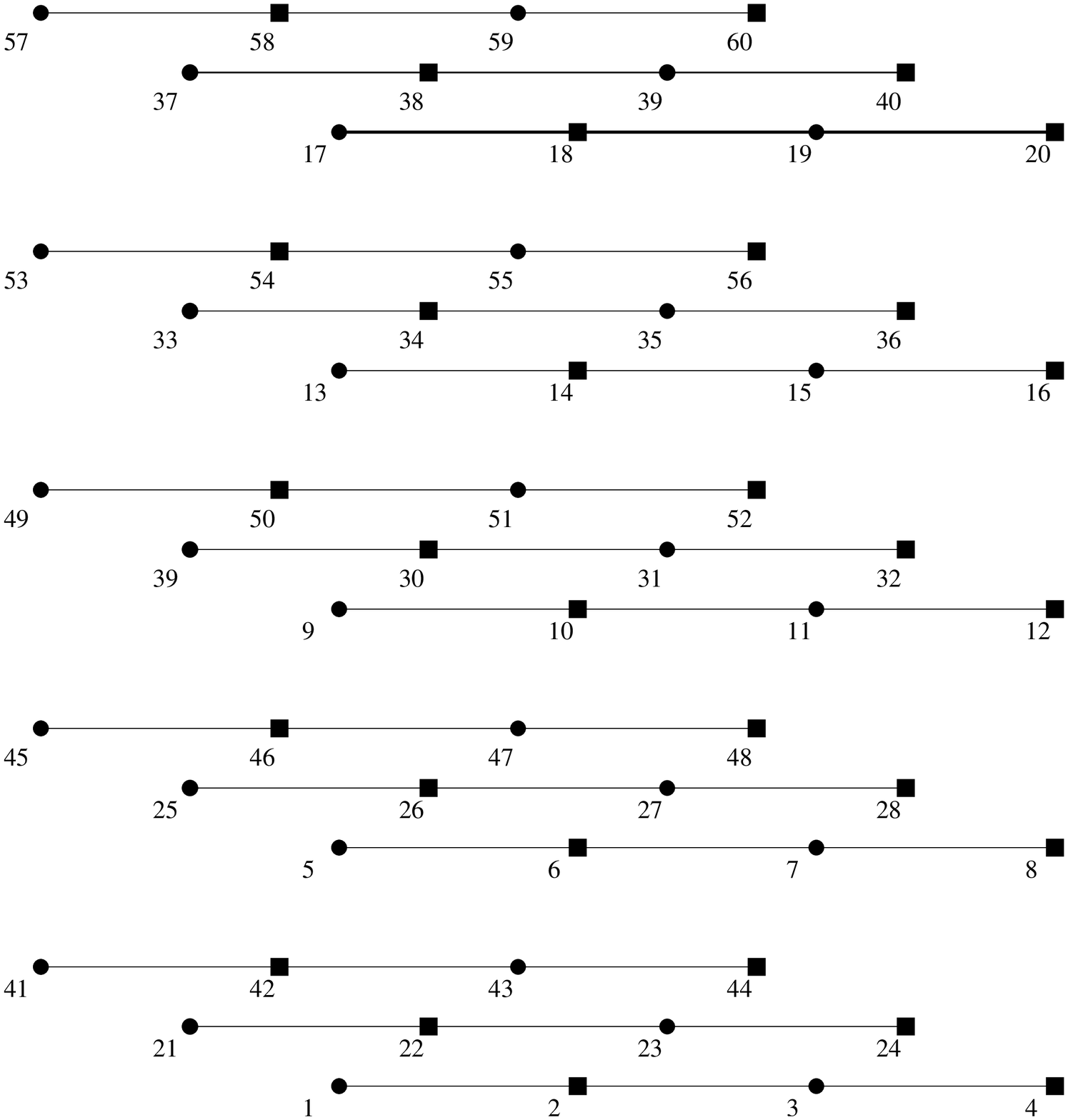}

\caption{\label{fig:3DExampleEven}  An example for the $(4,5,3)$-cluster
state and its resulting graph $G_{AA^c}$ between $A$ and its complement as
considered in Sec.~\ref{E_S for
bipartite splits}.
Here the vertices in $A$ are depicted by small
boxes $\vrule height4pt width3pt depth0pt$ . }
\end{figure}

\end{widetext}

Figure~\ref{fig:TreeExample1} gives an example for a tree for which the Schmidt measure does not
coincide with the size of the smaller bipartition, the upper bound according to
Proposition~\ref{twoc}.

{\it Example 2: The Schmidt measure of a 1D-, 2D-, and 3D-cluster state is
\begin{equation}
    E_{S}(|G\rangle) = \lfloor
    \frac{|V|}{2} \rfloor.
\end{equation}
}

{\it Proof:}
To see this, we only consider the $3D$ case, since the
former can be reduced to this. Moreover note that the $3D$-cluster
does not contain any (induced) cycles of odd length. Therefore,
it is 2-colorable and because of Eq.~(\ref{E for 2-colorable graphs}), we only
have to provide a bipartite
split with Schmidt rank $\lfloor  |V|/2 \rfloor$. For this
we choose a Cartesian numbering for the vertices starting in one
corner, i.e., $(x,y,z)$ with
$x=1,...,X$, $y=1,...,Y$ and  $z=1,...,Z$.

Let us first assume that $X$
is an even integer. Then let $A=\bigcup_{x\, \text{even}} A_x$ denote
the partition consisting
of vertices in planes $A_x$ with even $x$, and $y$ and $z$ being
unspecified. The graph
$G_{AA^c}$ consists of $Y\times Z$ parallel linear chains, which
alternately cross $A$ and $A^c$ (see Fig.\ \ref{fig:3DExampleEven}).
Since $|A|= (X/2) \times  Y \times Z $, we have to show that
for no subset $A'\subseteq A\;$ Eq.~(\ref{dependent neighborhoods}) holds.
This easily can be done, inductively showing, that
vertices in $A_x$ cannot be be contained in $A'$ for all even
$x=2,...,X$,
if  Eq.~(\ref{dependent neighborhoods})
shall be satisfied.

For $x=2$ this holds, because for every $a \in A'\cap A_2$ there is a
unique adjacent leaf $b \in A'\cap A_1$. Moreover, since
$b$ is a leaf, $n^b_a=1$ can only hold for one $a \in A'$. Therefore,
\begin{equation}
\sum_{a \in A'} n_a^b \not=_{{\mathbbm F}_2} 0.
\end{equation}
For even $x\geq 2$ note that, because $G$ is a tree, any two
$a_1,a_2 \in A_x$ have disjoint neighborhoods in $A_{x-1}$, i.e.,
\begin{equation}
    N_{a_1}\cap N_{a_2}\cap A_{x-1}=\emptyset.
\end{equation}
In order to fulfill Eq.~(\ref{dependent neighborhoods}),
any occurrence of $a \in A'\cap
A_x$ can therefore only be
compensated by some $a' \in A_{x-2}$, which is impossible by the
inductive presumption.

\begin{widetext}

\begin{figure}[th]

\includegraphics[width=14cm]{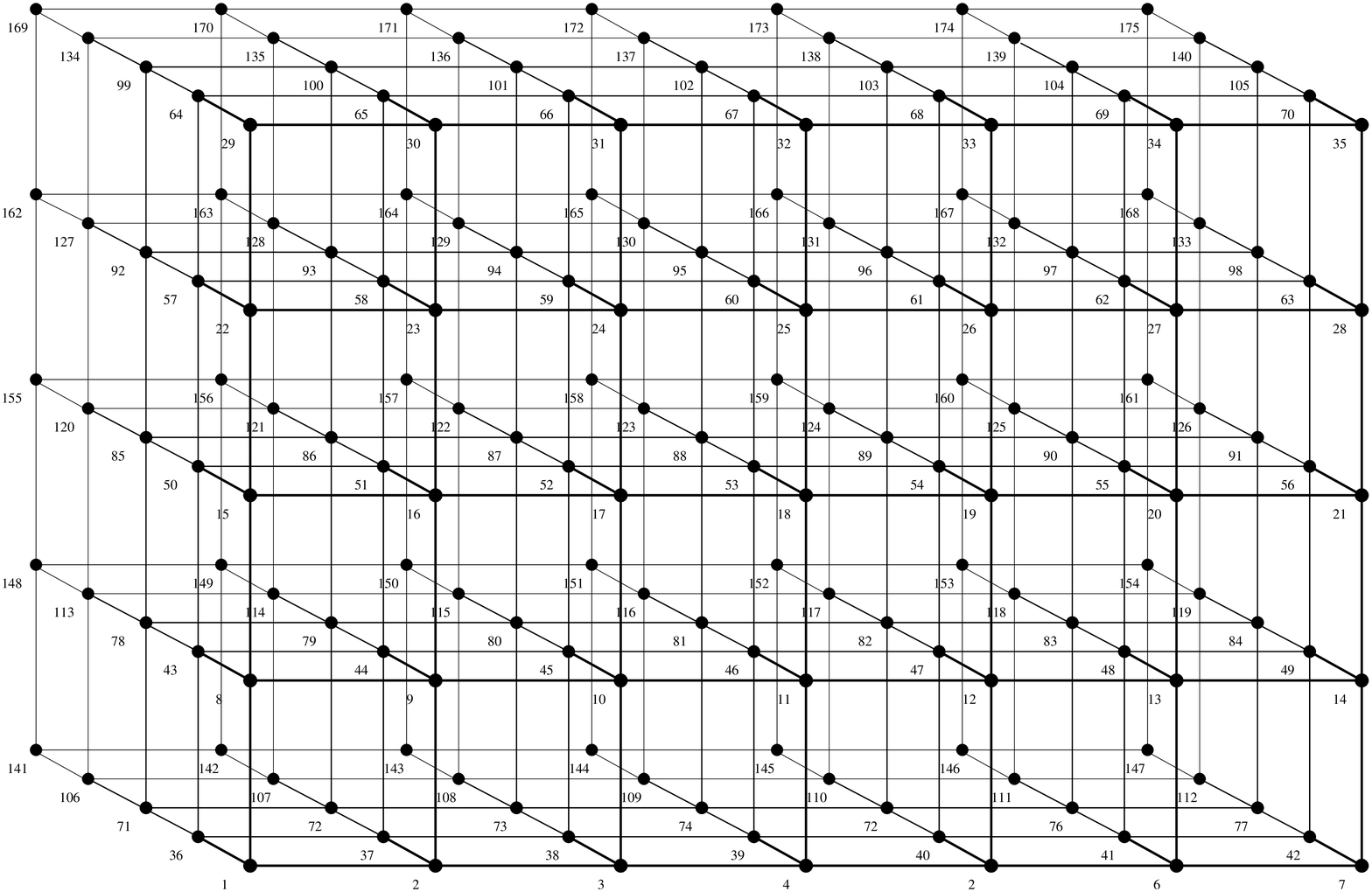}\

\vspace{.5cm}
\includegraphics[width=14cm]{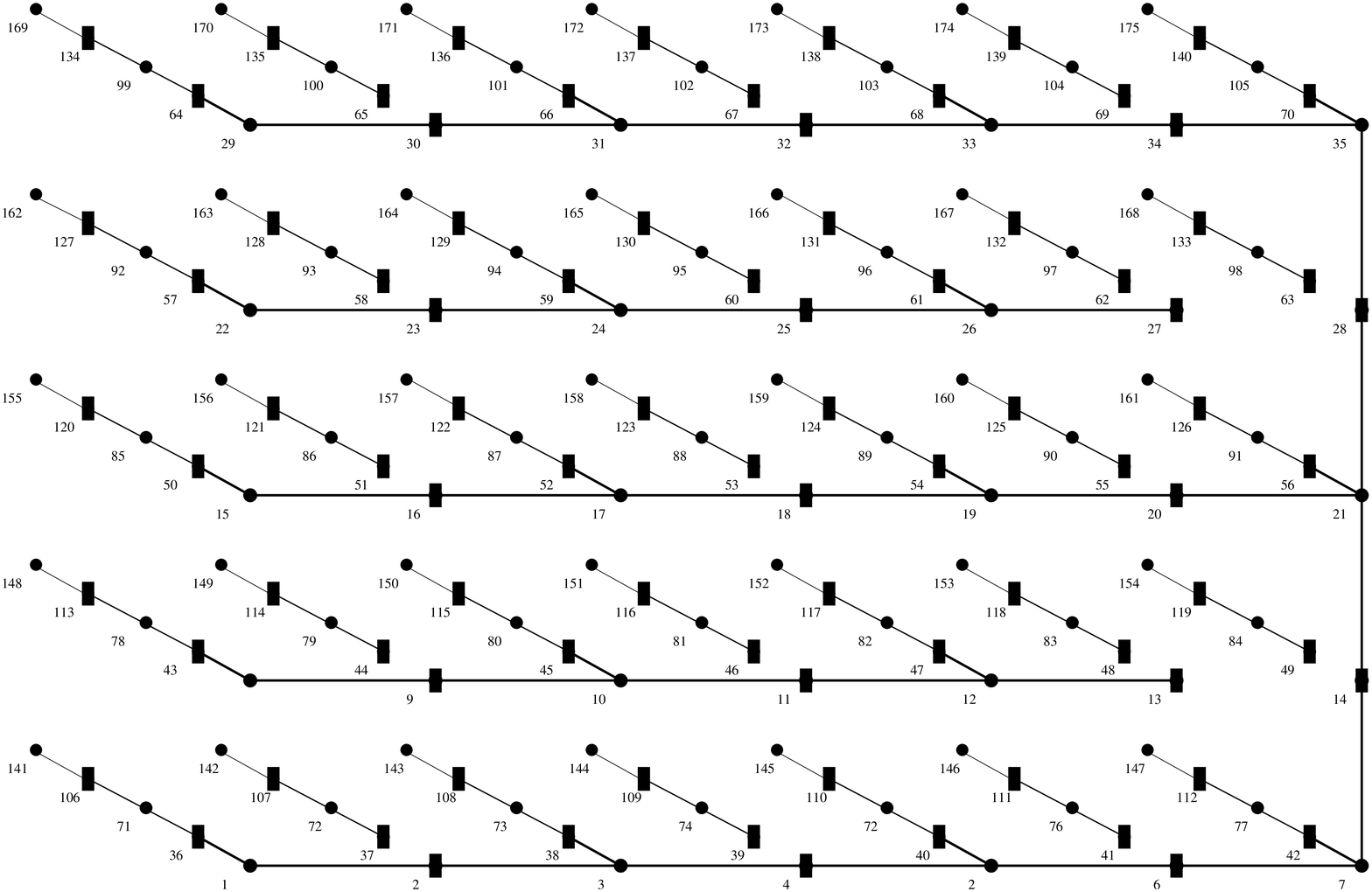}

\caption{\label{fig:3DExampleOdd}  An example for the 
$(7,5,5)$-cluster state and its resulting
graph $G_{AA^c}$
between $A$ and its complement as considered in Sec.~\ref{E_S for bipartite splits}.
Here the vertices in $A$ are depicted by small
boxes $\vrule height4pt width3pt depth0pt$.
The picture gives a rotated view on the cluster
considered in the
proof for the case, that $X$, $Y$, and $Z$ are odd integers:
The front plane, consisting of
the vertices $1$ until $35$, is the $y$-$z$-plane $A_X$ in the proof.   }
\end{figure}

\end{widetext}

In the case where $X$, $Y$ as well as $Z$ are odd integers, the
previous construction will yield a graph $G_{AA^c}$ consisting of
separate linear chains on
\begin{equation}
A=\bigcup_{x=1,...,X-1} A_x
\end{equation}
ending in the plane $A_X$ (see Fig.\ \ref{fig:3DExampleOdd}).
In this case we add every
second row $A_{Xy}$, $y=2,...,Z-1$,
to the partition $A$ as well as
of the last row $A_{XZ}$ every second vertex, giving the
size
\begin{eqnarray}
|A| =
\lfloor \frac{X}{2}\rfloor \times  Y \times Z \,+\,  \lfloor
\frac{Y\times Z}{2}\rfloor
=  \lfloor \frac{X \times Y\times Z }{2} \rfloor .
\end{eqnarray}
The inductive argument from above now still holds for all vertices in
$A$, except from the $y$-$z$-plane $A_x$ and can be continued
by a similar argument now considering the rows $A_{Xy}$ instead of
planes.
Note that the results could as well be obtained by simply applying
the sufficient
criterion in Proposition \ref{sufficient crit for max rank} to the
stated bipartitioning $(A,B)$. However, this inductive proof may be
of interest also for other graph classes.
\proofend

{\it Example 3: The Schmidt measure of
an entangled ring with an even number $|V|$ of
vertices is given by $|V|/2$.}

\begin{figure}[th]
\includegraphics[width=9cm]{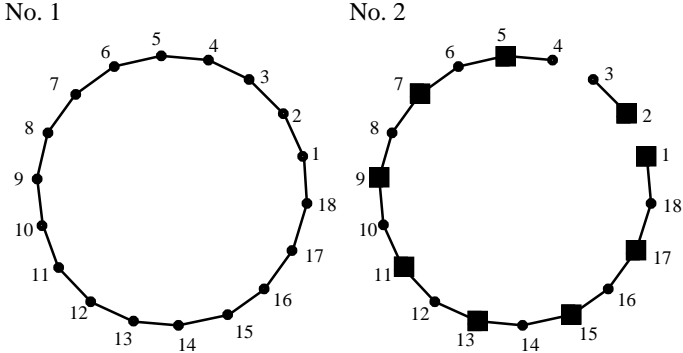}
\caption{\label{fig:EvenRingExample} Graph No.~1 is an entangled ring on $18$ vertices.
Graph No.~2 represents the resulting graph
between the partitions $A$, whose vertices are depicted by boxes, and the
partition $B$, whose vertices are depicted by discs.}
\end{figure}
{\it Proof:}
This is a 2-colorable graph, which gives on the one hand
the upper bound of $|V|/2$
for the Schmidt measure. On the other hand, by choosing the
partitions $A=\{1,2\}$ and $B=\{3,4\}$ on the first four vertices,
which are increased (for $|V|>4$)
alternately by the rest of the vertices, yielding the partitioning
with
\begin{eqnarray}
A&=&\{1,2,5,7,...,2k+5,...,|V|-1\}\\
B&=& \{3,4,6,8,...,2k+6,...,|V|\},
\end{eqnarray}
one obtains a bipartitioning $(A,B)$, which has maximal Schmidt rank $E_S^{(A,B)}=|V|/2$ according
to Proposition~\ref{sufficient crit for max rank} (see Fig.~\ref{fig:EvenRingExample}).
\proofend

{\it Example 4:
All connected graphs up to seven vertices.}
\begin{table}

\begin{tabular}{|cccccccc|}
\hline\hline
No.\ & $|\text{LUclass}|$ & $|V|$ & $|E|$ & $E_S$ &  $RI_3$ & $RI_2$ & $2-col$ \\
\hline\hline
1   & 1        & 2     &  1    &     1 &   &   & yes \\
2   & 2        & 3     &  2    &     1 &   &   & yes \\
3   & 2        & 4     &  3    &     1 & & (0,3)  & yes \\
4   & 4        & 4     &  3    &     2 & & (2,1)  & yes \\
5   & 2        & 4     &  4    &     1 & & (0,10)  & yes \\
6   & 6        & 5     &  4    &     2 & & (6,4)  & yes \\
7   & 10       & 5     &  4    &     2 & & (8,2)  & yes \\
8   & 3        & 5     &  5    & $2<3$ & & (10,0)  & no \\
9   & 2        & 6     &  5    &     1 & (0,0,10)& (0,15)  & yes \\
10  & 6        & 6     &  5    &     2 & (0,6,4)& (8,7)  & yes \\
11  & 4        & 6     &  5    &     2 & (0,9,1)& (8,7)  & yes \\
12  & 16       & 6     &  5    &     2 & (0,9,1)& (11,4)  & yes \\
13  & 10       & 6     &  5    &     3 & (4,4,2)& (12,3)  & yes \\
14  & 25       & 6     &  5    &     3 & (4,5,1)& (13,2)   & yes \\
15  & 5        & 6     &  6    &     2 & (0,10,0)& (12,3)   & yes \\
16  & 5        & 6     &  6    &     3 & (4,6,0)& (12,3)   & yes \\
17  & 21       & 6     &  6    &     3 & (4,6,0)& (14,1)   & yes \\
18  & 16       & 6     &  6    &     3 & (6,4,0)& (15,0)   & yes \\
19  & 2        & 6     &  9    &   $3<4$ & (10,0,0)& (15,0)   & no \\
20  & 2        & 7     &  6    &     1 & (0,0,35)& (0,21)   & yes \\
21  & 6        & 7     &  6    &     2 & (0,20,15)& (10,11)   & yes \\
22  & 6        & 7     &  6    &     2 & (0,30,5)& (12,9)   & yes \\
23  & 16       & 7     &  6    &     2 & (0,30,5)& (14,7)   & yes \\
24  & 10       & 7     &  6    &     2 & (0,33,2)& (15,6)   & yes \\
25  & 10       & 7     &  6    &     3 & (12,16,7)& (16,5)   & yes \\
26  & 16       & 7     &  6    &     3 & (12,20,3)& (16,5)   & yes \\
27  & 44       & 7     &  6    &     3 & (12,21,2)& (17,4)   & yes \\
28  & 44       & 7     &  6    &     3 & (16,16,3)& (18,3)   & yes \\
29  & 14       & 7     &  6    &     3 & (20,12,3)& (18,3)   & yes \\
30  & 66       & 7     &  6    &     3 & (20,13,2)& (19,2)  & yes \\
31  & 10       & 7     &  7    &     2 & (0,34,1)& (16,5)   & yes \\
32  & 10       & 7     &  7    &     3 & (12,22,1)& (16,5)   & no \\
33  & 21       & 7     &  7    &     3 & (12,22,1)& (18,3)   & no \\
34  & 26       & 7     &  7    &     3 & (16,18,1)& (18,3)   & yes \\
35  & 36       & 7     &  7    &    3  & (16,19,0)& (19,2)   & no \\
36  & 28       & 7     &  7    &     3 & (20,14,1)& (18,3)   & no \\
37  & 72       & 7     &  7    &     3 & (20,15,0)& (19,2)   & no \\
38  & 114      & 7     &  7    &   3   & (22,13,0)& (20,1)   & yes \\
39  & 56       & 7     &  7    &   $3<4$ & (24,10,1)& (20,1)   & no \\
40  & 92       & 7     &  7    &   $3<4$ & (28,7,0)& (21,0)   & no \\
41  & 57       & 7     &  8    &   $3<4$ & (26,9,0)& (20,1)   & no \\
42  & 33       & 7     &  8    &   $3<4$ & (28,7,0)& (21,0)   & no \\
43  & 9        & 7     &  9    &     3 & (28,7,0)& (21,0)   & yes \\
44  & 46       & 7     &  9    &   $3<4$ & (32,3,0)& (21,0)   & no \\
45  & 9        & 7     &  10   &   $3<4$ & (30,5,0)& (20,1)   & no \\
\hline\hline
\end{tabular}
\caption{\label{tab1} The number of vertices $|V|$ and edges $|E|$,
Schmidt measure $E_S$, rank index $RI_3$ and $RI_2$ (for splits with
2 or 3 vertices in the smaller partition), number
of non-isomorphic but LU equivalent graphs $|\text{LUclass}|$, and the
2-colorable property $2-col$ for the graph classes in Fig.
\ref{fig:List1} and \ref{fig:List2} .}

\end{table}

We have computed the lower and upper bounds to the Schmidt
measure, the Pauli persistency, and the maximal partial rank, for the
non equivalent graphs in Figs.~\ref{fig:List1} and \ref{fig:List2}.
They are listed in Table~\ref{tab1},
where we have also included the {\it rank index}. By the rank index, we simply
compressed the information contained in the Schmidt rank list with
respect to all bipartite splittings, counting
how many times a certain rank occurs in splittings with either two or
three vertices in the smaller partition. For example, the rank index
$RI_3=(20,12,3)$ of graph number $29$ means that the
rank $3$ occurs 20 times in all possible $3$-$4$-splits, the rank
$2$ twelve times, and the rank $1$ only three times. (Note, that here we use $\text{log}_2$ of the actual Schmidt rank.)
Similarly, because of $RI_2=(18,3)$ the rank $2$ ($1$) occurs $18$
($3$) times in  all $2$-$5$-splits of
the graph number $29$.

For connected graphs the Schmidt rank $0$ cannot occur for
any bipartite splitting $(A,B)$, since this would correspond to an
empty graph $G_{AB}$. Because the rank index is invariant under
permutations of
the partitions according to graph isomorphisms it provides information
about whether two graph states are equivalent under local unitaries
{\it plus} graph isomorphisms as
treated in Sec.~\ref{LU classes}. But note
that graph number $40$,
$42$ and $44$ are examples for non-equivalent graphs with the same
rank index.
Nevertheless, comparing the list of Schmidt ranks with respect to all
bipartitions in detail shows that no
permutation of the vertex set exists (especially none which is induced
by a proper graph isomorphism on both sides), which would cause a
permutation of the corresponding rank list, such that
two of the graphs could be locally equivalent.
In Table
\ref{tab1}
we have also listed the sizes of the corresponding equivalence classes
under LU and graph isomorphisms, as well as whether
$2$-colorable representatives exist.
For $295$ of $995$ non-isomorphic graphs the lower and
upper bound differs and that in these cases the Schmidt measure also
 non-integer values in
$\text{log}_2\{1,...,2^{|V|} \}$ are possible.
As has been discussed in Sec.~\ref{Schmidt measure},
in this paper we omit the computation of the
exact value for the Schmidt measure.

Moreover note that only graph number $8$ and $19$ have maximal
partial rank with respect to all bipartite splits. Entanglement here
is distributed symmetrically between all parties, which makes it
"difficult" to disentangle the state by few measurements.
From this one can understand why the gap between the lower and upper
bound occurs in such cases. As discussed in Sec.~\ref{E_S for bipartite splits}
of all graph codes with less than seven vertices only these two are candidates for strongly error detecting graph codes introduced in Ref.~\cite{schlinge01}.

\vspace{.5cm}

{\it Example 5: Concatenated $[7,1,3]$-CSS-code.}

\begin{figure}[th]
\includegraphics[width=8cm]{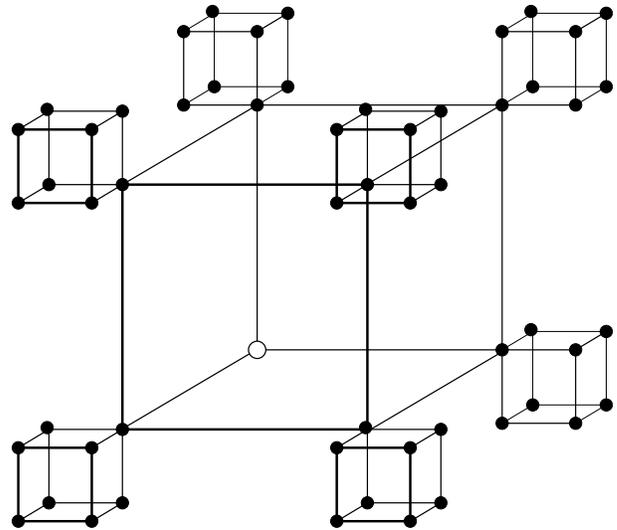}
\caption{\label{fig:bee} Resource graph state for the concatenated $[7,1,3]$-CSS code.}
\end{figure}

The graph $G$ depicted in Fig.\ref{fig:bee} represents an
encoding procedure for the concatenated $[7,1,3]$-CSS-code. The
corresponding graph state has Schmidt measure $28$. For encoding,
the qubit at the vertex $\circ$ can be in an arbitrary state. With
the rest of the vertices (initially prepared in the state
corresponding to $|x,+\rangle$), it is then entangled by the
$2$-qubit unitary $U^{(a,b)}$, introduced in Eq.~(\ref{IsingInteraction}). Encoding the state at vertex $\circ$
then means to perform $\sigma_x$-measurements at all vertices of
the inner square, yielding the corresponding encoded state on the
$7^2=49$ ``outer" vertices. The encoding procedure may
alternatively be realized by teleporting the bare qubit, initially
located on some ancillary particle, into the graph by performing a
Bell measurement on the ancilla and the vertex $\circ$ of the
graph state vector $|G'\rangle$. Here $|G'\rangle$ denotes the
graph state vector obtained from $|G\rangle$ by seven
$\sigma_x$-measurements at all vertices of the inner square except
$\circ$. In this sense $G'$ represents the resource for the
alternative  encoding procedure. It has maximal Schmidt measure
$25$, whereas the corresponding $0$ and $1$ code words have
Schmidt measure $24$. They can be obtained with probability $1/2$
from $|G'\rangle$ by a $\sigma_z$-measurement at the vertex
$\circ$.

\vspace{.5cm}

{\it Example 6: Quantum Fourier Transform (QFT) on three qubits.}

\begin{widetext}

\begin{figure}[th]

\includegraphics[width=8.5cm]{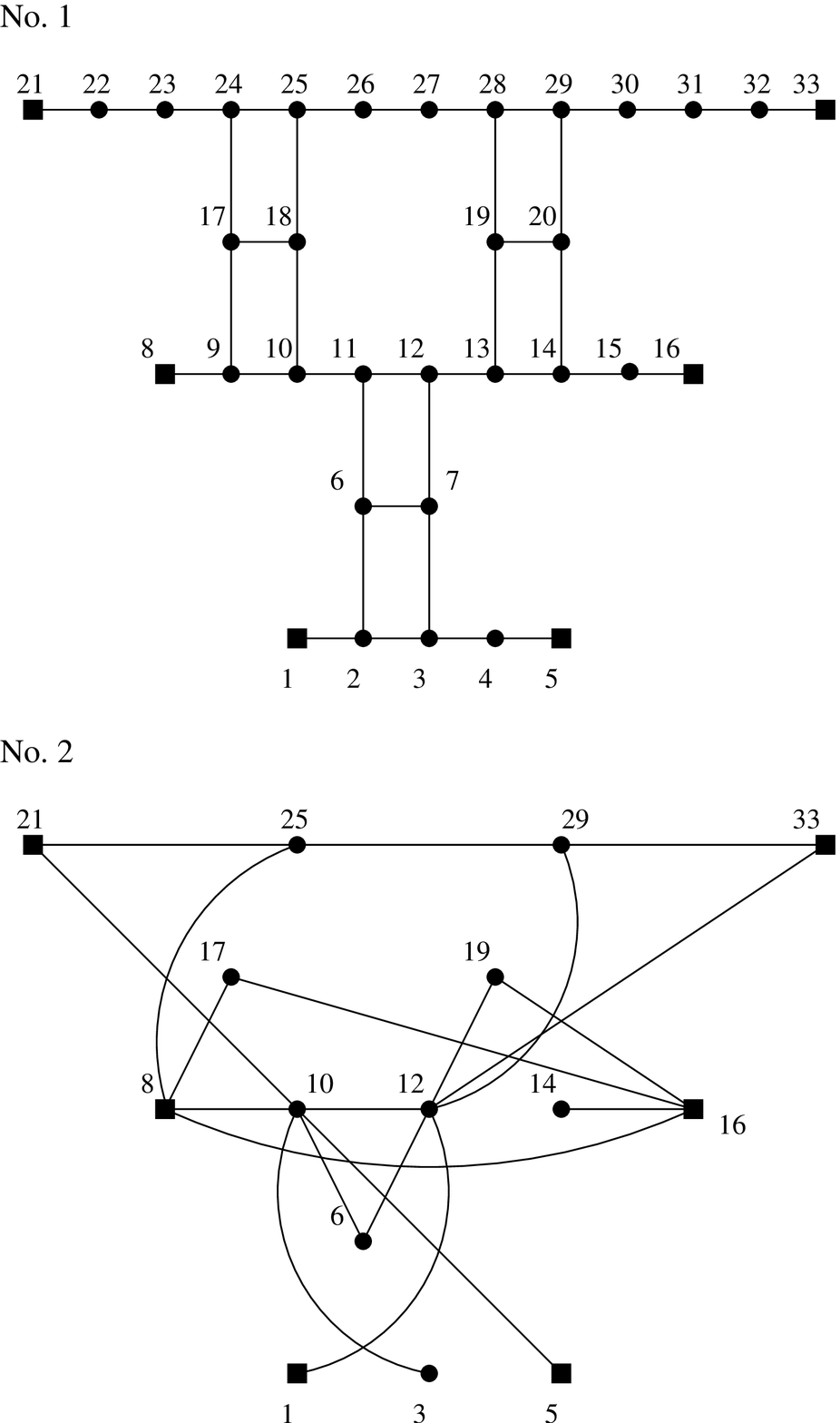}

\caption{\label{fig:3QFT} The graph associated with the QFT on 3
qubits in the one-way quantum computer is represented in graph
No.~1, where the boxes denote the input (left) and output (right)
vertices. Graph No.~2 is obtained from the first after performing
all Pauli measurements according to the protocol in Ref.~\cite{raussen03}, except from the $\sigma_x$-measurements at the
input vertices. More precisely, it is obtained from graph No.~1
after $\sigma_y$-measurements on the vertices $22, 23, 24, 26, 27,
28, 30, 31, 32$ and $\sigma_x$-measurements on the vertices $2, 4,
7, 9, 11, 13, 15, 18, 20$ have been performed.
}
\end{figure}

\end{widetext}

The graph No.~1 in Fig.~\ref{fig:3QFT} is a
simple example of an entangled graph state as it occurs in the
one-way computer of Refs.~\cite{raussen03,OneWay}. This specific
example represents the initial resource (part of a cluster)
required for the quantum Fourier transform QFT on 3 qubits
\cite{raussen03}. It has Schmidt measure $15$, where the partition
\begin{eqnarray}
        A=\{2, 4, 7, 9, 11, 13, 15, 18, 20, 22, 24, 26, 28, 30, 32\}
        \nonumber\\
\end{eqnarray} 
is a minimal vertex cover with maximal Schmidt rank. In the process
of performing the QFT, all vertices except the output vertices
$5,16,33$ are measured locally. During this process, the
entanglement of the resource state (with respect to every
partitioning) can only decrease.  Similar as with the graph state vector
$|G'\rangle$ obtained from Fig.~\ref{fig:bee}, graph No.~2
represents the input-independent resource needed for the essential
(non-Clifford) part of the QFT protocol \cite{raussen03}. It has
Schmidt measure $5$, where the partition $A=\{2, 9, 10, 11, 15\}$
now provides a minimal vertex cover with maximal Schmidt rank.

\section{Summary, discussion, and outline of further work}

In this paper we have developed methods that allow for a
qualitative and quantitative description of the multi-particle
entanglement that one encounters in graph states. Such graph
states capture the intuition of an interaction pattern between
quantum systems, with important applications in quantum error
correction, quantum communication, and quantum computation in the
context of the one-way quantum computer. The Schmidt measure is
tailored for a comparably detailed account on the quantum
correlations grasping genuine multi-particle entanglement, yet it
turns out to be computable for many graph states. We have
presented a number of general rules that can be applied  when
approaching the problem of evaluating the Schmidt measure for
general graph states, which are stated mostly in graph theoretical
terms. These rules have then been applied to a number of graph
states that appear in quantum computation and error correction.
Also, all connected graphs with up to seven vertices have been
discussed in detail. The formalism that we present here abstracts
from the actual physical realisation, but as has been pointed out
in several instances, a number of well-controllable physical
systems such as neutral atoms in optical lattices serve as
potential candidates to realize such graph states \cite{E1,E2}.

In this paper, the Schmidt measure has been employed
to quantify the degree of entanglement, as a generalization of the
Schmidt rank in the bipartite setting. This measure is sufficiently coarse to
be accessible for systems consisting of many constituents and to
allow for an appropriate discussion of multi-particle entanglement in 
graph states. The approach of quantifying 
entanglement in terms of rates of asymptotic reversible state 
transformations, as an alternative, appears unfeasible in the
many-partite setting. 
The question of the minimal reversible entangling generating 
set (MREGS) in multipartite systems
remains unresolved to date, even for quantum systems consisting of three qubits, and despite
considerable research effort \cite{MREGS,MREGS2}.
These MREGS are the (not necessarily finite)
sets of those pure states from which any other pure states
can be asymptotically prepared in a reversible manner under local operations with classical
communication (LOCC).
Hence, it seems unrealistic to date
to expect to be able to characterize multi-particle entanglement by the rates that can be
achieved in reversible asymptotic state transformations, analogous to the entanglement
cost and the distillable entanglement
under LOCC operations in bipartite systems.
In turn, such a description, if it was to
be found, could well turn out to be too detailed to capture
entanglement as an algorithmic resource
in the context of error correction or the one-way quantum computer, where, needless to say,
distributed quantum systems with very many constituents are encountered.

For future investigations, a more feasible
characterization of LU equivalence would open up further
possibilities. A step that would go significantly beyond the
treatment of the present paper would be to consider measurements
corresponding to observables not contained in the Pauli-group.
Unfortunately, in this case the stabilizer formalism is no longer
available, at least not in the way we used it in this paper. Such
an extension would, however, allow for a complete monitoring of the
entanglement resource as it is processed during a quantum
computation in the one-way computer, where also measurements in
tilted bases play a role.

Finally, taking a somewhat different perspective, one could also extend the
identification of edges with interactions to weighted graphs, where a real
positive number associated with each edge characterizes the interaction
strength (e.g., the interaction time).
With such a notion at hand, one could study the quantum correlations
as they emerge in more natural systems. One example is given by a
Boltzmann system of particles, where each particle follows a
classical trajectory but carries a quantum degree of freedom that is affected
whenever two particles scatter.
With techniques of random graphs, it would be interesting to investigate
what kind of multi-particle correlations are being built up when the system
starts from a prescribed initial state, or to study the steady state.
The answer to these questions depends on the knowledge of the
interaction history. A hypothetical observer who is aware of the
exact distribution in classical phase space (Laplacian
demon perspective) would assign a definite graph corresponding to a
pure entangled state to the ensemble. An observer who lacks all or part
of this classical information about the particles' trajectories
would describe the state by a random mixture of graphs and corresponding
quantum states. One example of this latter situation would be a
Maxwell demon scenario in which one studies the bipartite entanglement
as it builds up between two parts of a container.

\section{Acknowledgements} We would like to acknowledge fruitful
discussions with D.\ Schlingemann, M.\ Van den Nest, P.\ Aliferis, 
as well as with H.\ Aschauer, W.\ D{\"u}r, and R.\ Raussendorf. For valuable hints 
on connections to known results in graph theory \cite{Bouchet} and multi-linear algebra
\cite{Lathauwer} we would like to thank G.\ Royle, K.\ Audenaert and the referee. 
This work has been supported by the Deutsche Forschungsgemeinschaft
(Schwerpunkt QIV), the Alexander von Humboldt Foundation (Feodor
Lynen Grant of JE), the European Commission (IST-2001-38877/-39227, IST-1999-11053),
and the European Science Foundation.


\begin{thebibliography}{99}


\bibitem{Graph}
D.B.\ West, {\it Introduction to Graph Theory} (Prentice Hall, Upper
Saddle River, 2001).

\bibitem{Graph2}
R.\ Diestel, {\it Graph Theory} (Springer, Heidelberg, 2000).

\bibitem{raussen03}
R.\ Raussendorf, D.E.\ Browne, and H.J.\ Briegel,
Phys.\ Rev.\ A {\bf 68}, 022312 (2003).

\bibitem{schlinge03}
D.\ Schlingemann, Quant.\ Inf.\ Comp.\ {\bf 2}, 307 (2002);
Quant. Inf. Comp. {\bf 4}, No 4, 287-324 (2004).

\bibitem{Wolfgang}
W.\ D{\"u}r, H.\ Aschauer, and H.J.\ Briegel, Phys.\ Rev.\ Lett.\ {\bf 91}, 107903 (2003).

\bibitem{Grassl}
M.\ Grassl, A.\ Klappenecker, and M.\ R{\"o}tteler,
{\it Graphs, Quadratic Forms, and
Quantum Codes}, in Proc.\ 2002 IEEE International Symposium on
Information Theory, Lausanne, Switzerland (2002), page 45.

\bibitem{schlinge01}
D.\ Schlingemann and R.F.\ Werner,  Phys.\ Rev.\ A {\bf 65}, 012308
(2002).

\bibitem{Gottesman}
D.\ Gottesman, {\em Stabilizer Codes and Quantum Error Correction},  PhD thesis
(CalTech, Pasadena, 1997).

\bibitem{Cluster}
H.J.\ Briegel and R.\ Raussendorf, Phys.\ Rev.\ Lett.\ {\bf 86}, 910 (2001).

\bibitem{OneWay}
R.\ Raussendorf and H.J.\ Briegel, Phys.\ Rev.\ Lett.\ {\bf 86},  5188 (2001).

\bibitem{Schmidt}
J.\ Eisert and H.J.\ Briegel, Phys.\ Rev.\ A {\bf 64}, 022306  (2001).

\bibitem{Nielsen}
T.J.\ Osborne and M.A.\ Nielsen, Phys.\ Rev.\ A {\bf 66}, 032110 (2002).

\bibitem{Osterloh}
A.\ Osterloh, L.\ Amico, G.\ Falci, and R.\ Fazio, Nature {\bf 416}, 608 (2002).

\bibitem{Latorre}
J.I.\ Latorre, E.\ Rico, and G.\ Vidal, Quantum Inf.\ Comput.\ {\bf 4}, 048 (2004).

\bibitem{Auden}
K.\ Audenaert, J.\ Eisert, M.B.\ Plenio, and R.F.\ Werner, Phys.\ Rev.\ A {\bf 66}, 042327 (2002).

\bibitem{Wolf}
M.M.\ Wolf, F.\ Verstraete, and J.I.\ Cirac, Phys.\ Rev.\ Lett.\ {\bf 92}, 087903 (2004).

\bibitem{Hideo}
J.K.\ Stockton, J.M.\ Geremia, A.C.\ Doherty, and H.\ Mabuchi, Phys.\ Rev.\ A {\bf 67}, 022112 (2003).

\bibitem{Rings}
K.M.\ O'Connor and W.K.\ Wootters, Phys.\ Rev.\ A {\bf 63}, 052302 (2001).

\bibitem{Frank}
F.\ Verstraete, M.\ Popp, and J.I.\ Cirac,  Phys.\ Rev.\ Lett.\  {\bf 92}, 027901 (2004).

\bibitem{Buzek}
M.\ Plesch and V.\ Buzek, Phys.\ Rev.\ A {\bf 67}, 012322 (2003); Phys.\ Rev.\ A {\bf 68}, 012313 (2003).

\bibitem{Zanardi}
P.\ Giorda and P.\ Zanardi, Phys.\ Rev.\ A {\bf 68}, 062108 (2003).
  
\bibitem{Parker}
M.G.\ Parker and V.\ Rijmen, {\it The Quantum Entanglement of Binary and Bipolar Sequences}, Sequences and Their Applications, SETA'01, Discrete Mathematics and Theoretical Computer Science Series, Springer, 2001, Ed.: T.Helleseth, P.V.Kumar and K.Yang.


\bibitem{Tangle}
V.\ Coffman, J.\ Kundu, and W.K.\ Wootters, Phys.\ Rev.\ A {\bf 61},  052306 (2000).

\bibitem{Plenio}
M.B.\ Plenio and V.\ Vedral, J.\ Phys.\ A {\bf 34}, 6997  (2001).

\bibitem{Meyer}
D.\ Meyer and N.\ Wallach, J.\ Math.\ Phys.\ {\bf 43}, 4273 (2002).

\bibitem{Wei}
T.-C.\ Wei and P.M.\ Goldbart, Phys.\ Rev.\ A {\bf 68}, 042307 (2003).

\bibitem{Barnum}
H.\ Barnum and N.\ Linden, J.\ Phys.\ A  {\bf 34}, 6787  (2001).

\bibitem{Guifre}
W.\ D{\"u}r, G.\ Vidal, and J.I.\ Cirac, Phys.\ Rev.\ A {\bf 62}, 062314 (2000).

\bibitem{Lathauwer}
For a more detailed discussion of numerical issues we refer to: 
L.\ De Lathauwer, {\em Signal Processing based on Multilinear Algebra},  PhD thesis
(Dept.\ of Electrical Engineering ESAT, K.U.\ Leuven, 1997), ESAT-SISTA/TR 1997-74,
where the problem of finding the minimal linear decomposition of a pure state (rank-$N$ tensors) into  product states (product of rank-1 tensors) is discussed in the context of the 'Canonical Decomposition' or  'Parallel Factors' problem.

\bibitem{footnote-to-KGT}
This is the content of the so-called Gottesman-Knill theorem as it is stated in Ref.~\cite{NielsenBook} (Theorem 10.7).

\bibitem{schlinge02}
D.\ Schlingemann, Quant.\ Inf.\ Comp.\ {\bf 2}, 307 (2002).

\bibitem{SimulClassical}
G.\ Vidal,  Phys.\ Rev.\ Lett.\ {\bf 91}, 147902 (2003).


\bibitem{Classification}
Similar classifications for self-dual additive codes over $GF(4)$ have been provided independently, e.g., for $N\leq 7$ in: G. H\"ohn, Mathematische Annalen {\bf 327}, pp. 227-255 (2003), for $N\leq 9$ in: D.G. Glynn, T.A. Gulliver, J.G. Maks, and M. K. Gupta, {\it The geometry of additive quantum codes}, submitted to Springer-Verlag (2004), and for $N\leq 12$ in: L. E. Danielsen and M.G. Parker, E-print math.CO/0504522.  


\bibitem{Bouchet}
A.\ Bouchet, Discrete Math.\ {\bf 114}, 75  (1993).

\bibitem{Maarten}
M.\ Van den Nest, J.\ Dehaene, and B.\ De Moor, Phys.\ Rev.\ A {\bf 69}, 022316 (2004);  Phys. Rev. A {\bf 70}, 034302 (2004).


\bibitem{Glynn02}
D.G. Glynn, {\it On self-dual quantum codes and graphs}, Submitted to Electronic Journal of Combinatorics,  Preprint (2002).


\bibitem{BouchetAlternative}
An alternative proof in terms of isotropic systems has already been provided in \cite{Bouchet}.

\bibitem{Flaas}
This counter example is due to: D. Fon-Der-Flaass, {\it Local equivalence of transversals in matroids}, Electr. J. Comb. {\bf 3}(1) (1996).

\bibitem{Hans}
H. Aschauer, J. Calsamiglia, M. Hein and H.J. Briegel,  Quant. Inf. Comp. {\bf 4}, 383 (2004).

\bibitem{MaartenInv}
M.\ Van den Nest, J.\ Dehaene, and B.\ De Moor, Phys. Rev. A {\bf 71}, 022310 (2005); E-print quant-ph/0410165;  E-print quant-ph/0411115.


\bibitem{NielsenBook}
M.A.\ Nielsen and I.L.\ Chuang, {\it Quantum Computation and Information} (Cambridge University Press, Cambridge, 2000).

\bibitem{Simulate0}
D.\ Gottesman, {\it The Heisenberg Representation of Quantum
Computers}, in Proceedings of the XXII International Colloquium on Group Theoretical Methods
in Physics, eds.\ S.P.\ Corney et al, (Cambridge, MA, International Press, 1999).

\bibitem{Simulate}
J.\ Eisert, K.\ Jacobs, P.\ Papadopoulos, and M.B.\ Plenio, Phys.\ Rev.\ A {\bf 62}, 052317 (2000).

\bibitem{Simulate1}
D.\ Collins, N.\ Linden, and S.\ Popescu, Phys.\ Rev.\ A {\bf 64}, 032302 (2001).


\bibitem{E1}
D.\ Jaksch, H.-J.\ Briegel, J.I.\ Cirac, C.W.\ Gardiner, and P.\ Zoller,
Phys.\ Rev.\ Lett.\ {\bf 82}, 1975  (1999).

\bibitem{E2}
L.-M.\ Duan, E.\ Demler, and M.D.\ Lukin, Phys.\ Rev.\ Lett.\ {\bf 91}, 090402 (2003).

\bibitem{MREGS}
C.H.\ Bennett, S.\ Popescu, D.\ Rohrlich, J.A.\ Smolin, and A.V.\ Thapliyal,
Phys.\ Rev.\ A {\bf 63}, 012307 (2001).

\bibitem{MREGS2}
N.\ Linden, S.\ Popescu, B.\ Schumacher, and 
M.\ Westmoreland, quant-ph/9912039; E.F.\ Galvao, M.B.\ Plenio, and
S.\ Virmani, J.\ Phys.\ A {\bf 33}, 8809 (2000); S.\ Wu and Y.\ Zhang,
Phys.\ Rev.\ A {\bf 63}, 012308 (2001); A.\ Acin, G.\ Vidal, and J.I.\ Cirac,
Quant.\ Inf.\ Comp.\ {\bf 3}, 55 (2003).

\end{thebibliography}
\end{document}